\documentclass[12pt,preprint]{aastex}

\newcommand{\kmpers}{{{\rm \: km\:s}^{-1}}}

\newcommand{\logNH}{{\log{{\rm(N/H)}}}}
\newcommand{\Htot}{{N{\rm(H_{tot})}}}

\newcommand{\NH}{{\rm N/H}}
\newcommand{\NHI}{{N{\rm(HI)}}}

\newcommand{\NNI}{{N{\rm(NI)}}}
\newcommand{\Hmol}{{\rm H_{2}}}
\newcommand{\Nmol}{{\rm N_{2}}}
\newcommand{\nHvol}{{\rm n_{H}}}
\newcommand{\nelec}{{\rm n_{e}}}
\newcommand{\NHmol}{{N\rm(H_{2})}}
\newcommand{\fHmol}{{f\rm(H_{2})}}
\newcommand{\logHI}{{\log{\NHI}}}
\newcommand{\logHmol}{{\log{\NHmol}}}
\newcommand{\logHtot}{{\log{\Htot}}}

\newcommand{\ppm}{{\rm \: ppm}}
\newcommand{\dex}{{\rm \: dex}}
\newcommand{\magnitude}{{\rm \: mag}}

\received{}
\revised{}
\accepted{11 September 2006}

\shorttitle{}

\shortauthors{Jensen, Rachford, and Snow}

\begin{document}

\title{Is There Enhanced Depletion of Gas-Phase Nitrogen in Moderately Reddened Lines of Sight?}

\author{Adam G. Jensen\altaffilmark{1}, Brian L. Rachford\altaffilmark{1,2}, and Theodore P. Snow\altaffilmark{1}}

\altaffiltext{1}{Center for Astrophysics and Space Astronomy, University of Colorado at Boulder, Campus Box 389, Boulder, CO 80309-0389; Adam.Jensen@colorado.edu, tsnow@casa.colorado.edu}
\altaffiltext{2}{Current affiliation:  Embry-Riddle Aeronautical University, 3700 Willow Creek Road, Prescott, AZ 86031-3720; Brian.Rachford@erau.edu}


\begin{abstract}

We report on the abundance of interstellar neutral nitrogen (NI) for 30 sightlines, using data from the {\it Far Ultraviolet Spectroscopic Explorer} ({\it FUSE}) and the {\it Hubble Space Telescope} ({\it HST}).  NI column densities are derived by measuring the equivalent widths of several ultraviolet absorption lines and subsequently fitting those to a curve of growth.  We find a mean interstellar N/H of $51\pm4\ppm$.  This is below the mean found by \citeauthor{Meyer} of $62^{+4}_{-3}\ppm$ (adjusted for a difference in $f$-values).  Our mean N/H is similar, however, to the ($f$-value adjusted) mean of $51\pm3\ppm$ found by \citeauthor{Knauth} for a larger sample of sightlines with larger hydrogen column densities comparable to those in this study.  We discuss the question of whether or not nitrogen shows increased gas-phase depletion in lines of sight with column densities $\logHtot \gtrsim21$, as claimed by \citeauthor{Knauth}.  The nitrogen abundance in the line of sight toward HD 152236 is particularly interesting.  We derive very small N/H and N/O ratios for this line of sight that may support a previous suggestion that members of the Sco OB1 association formed from an N-deficient region.

\end{abstract}


\section{INTRODUCTION AND BACKGROUND}
\label{s:intro}
Many atomic species in the interstellar medium (ISM) show enhanced depletions relative to hydrogen in dense, highly reddened environments.  For the most part, however, neutral nitrogen (NI) has appeared to be an exception.  Because nitrogen is not expected to be a significant source of material for dust grains \citep{SnowWitt}, and the abundances of molecules such as $\Nmol$ also appear to be small in the ISM \citep{Knauth}, it is not surprising that nitrogen depletions may be small.  Therefore, if enhanced nitrogen depletion in a given line of sight is detected, it would challenge current models of interstellar dust composition and/or nitrogen chemistry.

In diffuse interstellar clouds, which are permeated by a radiation field whose high-energy cut-off is 13.60 eV, nitrogen is primarily in the form of NI because its ionization potential of 14.53 eV is slightly higher than HI.  Hence, the analysis of nitrogen depletions depends on this assumption that nitrogen in HI regions remains largely unaffected by ionization and therefore the total nitrogen abundance relative to hydrogen, N/H, is accurately approximated by $\NNI$/[$\NHI + 2\NHmol$].

One of the earliest studies of interstellar NI, by \citet{Lugger}, attempted to empirically determine the $f$-values of many NI absorption lines in the {\it Copernicus} wavelength range.  The unknown $f$-values were determined from one line of sight where several lines with known oscillator strengths were on the linear portion of the curve of growth.  This new information was used to determine column densities for eight other sightlines.  Their study found varying depletions of nitrogen, but with large errors.  The determined abundances were all within a factor of four of the best solar value at that time \citep{Withbroe}.

\citet{Ferlet} observed many additional lines of sight with {\it Copernicus}.  Using the $f$-values determined by \citeauthor{Lugger}, \citeauthor{Ferlet} found the nitrogen abundance to be independent of reddening (i.e. total hydrogen content) in the line of sight.  \citeauthor{Ferlet} also found the average nitrogen abundance to be consistent with a solar abundance within the large uncertainties (the solar abundance has been subsequently revised; see discussion below).  Also using {\it Copernicus} observations, \citet*{Hibbert} found N/H to be mostly independent of reddening.  A small exception was that less reddened lines of sight appeared to have slightly more nitrogen depletion than reddened lines of sight.  \citeauthor{Hibbert} were unable to rule out the possibility that hydrogen column densities were overestimated in the unreddened cases.

In further {\it Copernicus} observations, \citet{York} determined the average value of N/H in the ISM to be 50\% of the solar value, and also noted it was independent of reddening.  While the \citeauthor{York} value for N/H in the ISM and the solar value of N/H could be reconciled somewhat within the uncertainties, their results were the first to suggest that nitrogen in the ISM must be somewhat depleted if the solar abundances of various elements are appropriate standards to apply to the ISM.  It is worth noting that the lines of sight considered by \citeauthor{York} had a visual extinction of 1 magnitude at the most.  Additionally, \citeauthor{York} used the $f$-values presented by \citet{Cowan}, which are much larger than the $f$-values determined by \citet{Lugger}.  Therefore, the nitrogen column densities of \citeauthor{York} were systematically smaller than the nitrogen column densities of prior studies.

\citet*{Meyer} used the Goddard High Resolution Spectrometer (GHRS) onboard the {\it Hubble Space Telescope} to study lines of sight with low visual extinction ($A_V \lesssim 1 \magnitude$).  Their study found the value of N/H in the ISM to be 80\% of the \citet{Grevesse} solar value ($75\pm4\ppm$ for the ISM compared to $93\pm16\ppm$).  However, within the uncertainties, the results of \citeauthor{Meyer} could not rule out the possibility of either a solar N/H abundance or an abundance of N/H that is $2/3$ solar.  The case of a 2/3 solar N/H abundance would have been consistent with depletions observed at the time for O/H and Kr/H.  However, the idea that many ISM abundances are 2/3 of the solar abundance is no longer accepted \citep{SofiaMeyer}, based largely on revisions to the solar abundances that have brought ISM abundances such as O/H into closer but not necessarily complete agreement with the solar values \citep[see also the discussion and references in][]{Jensen}.  NI column densities determined by \citeauthor{Meyer} were based exclusively on measurements of the weak 1160 \AA{} doublet.  Therefore, it is important to note that the uncertainties of the \citeauthor{Meyer} study included these $f$-values of the weak doublet, determined by \citet{Hibbert}.  Adjusted for the more recent $f$-values calculated by \citet{Tachiev} and used in this paper, the \citeauthor{Meyer} mean is $62^{+4}_{-3}\ppm$ (see \S\ref{ss:1160doublet} for more discussion of these $f$-values).  For the sake of direct comparison, throughout the rest of this paper we quote this adjusted mean.

\citet{Moos} used the {\it Far Ultraviolet Spectroscopic Explorer} ({\it FUSE}) to study the H, D, N, and O content of several nearby sightlines and discovered variations in the N/H and N/O abundances.  \citeauthor{Moos} concluded that this difference is due to local ionization.

Recently, \citet{Knauth} measured NI column densities in lines of sight with $A_V \gtrsim 1 \magnitude$ using data from {\it FUSE} and the Space Telescope Imaging Spectrometer (STIS) onboard {\it HST}.  They found the first significant evidence of enhanced nitrogen depletion as total hydrogen column density increases.  Specifically, \citeauthor{Knauth} found nitrogen abundances less than the \citeauthor{Meyer} mean with respect to both hydrogen and oxygen for lines of sight with $\logHtot \gtrsim 21$.  \citeauthor{Knauth} also placed limits on the $\Nmol$ abundance that ruled it out as a significant sink of NI in the lines of sight with enhanced depletions.  The weighted average N/H abundance \citeauthor{Knauth} found, including lines of sight from \citeauthor{Meyer} (with some final N/H ratios slightly adjusted),\citet{Jenkins1999}, and \citet{Sonneborn}---the latter two sources using IMAPS and different absorption lines---is $59\pm2\ppm$; adjusting the \citeauthor{Knauth} and \citeauthor{Meyer} results for the \citet{Tachiev} $f$-values, this mean abundance is $54\pm2\ppm$.  The $f$-value adjusted weighted average of the 17 lines of sight analyzed by \citeauthor{Knauth} is $51\pm3$.  As with the \citeauthor{Meyer} mean, we quote this adjusted mean throughout the rest of this paper.

The purpose of this paper is to examine the NI content of many additional moderately reddened lines of sight, some of which have more $\Hmol$ and a higher $A_V$ than typical lines of sight in previous studies.

\section{OBSERVATIONS AND DATA REDUCTION}
\label{s:obsdata}
Data were taken from the {\it FUSE} and {\it HST} archives.  The sightlines were chosen from a sample observed by {\it FUSE} as part of a molecular hydrogen survey conducted by \citet{Rachford} and continued in B. L. Rachford et al. \citetext{in preparation}.  The original {\it FUSE} sample included 37 sightlines, and relevant {\it HST} data were available for eight of these sightlines.  For three sightlines (HD 166734, HD 281159, and NGC 2264 67, i.e. Walker 67) we could not reliably measure any NI absorption lines.  Additionally, we detected the 1134 \AA{} triplet of NI in the spectrum for the lines of sight toward HD 43384, HD 62542, HD 164740, and HD 210121.  However, the column densities that might be derived from profile fitting or a curve-of-growth method based on this moderately-strong triplet alone would be very uncertain because the profile is relatively insensitive to column density and $b$-value.  Systematic error might also be a concern because a variety of multiple-component velocity dispersions could conceivably replicate the profiles, but with a very different total column density than derived from a single-component profile fit or curve of growth.

Thus, our sample is reduced to 30 lines of sight where we can calculate a column density for NI, including 26 lines of sight for which we can reliably calculate N/H based on direct measurements of both HI and $\Hmol$.  In the remaining four cases, direct measurements of $\Hmol$ are available, but HI must be inferred.  In three of these remaining four cases we use the relationship presented by \citet{BohlinSavDrake} to infer the total hydrogen column density from $E_{B-V}$; in the fourth case we quote a literature value \citep{Hanson} where $\NHI$ is inferred from $E_{B-V}$ (see \S\ref{ss:hydrogen}).  Basic stellar data for the stars of these sightlines are given in Table \ref{stellardata}.  Values for hydrogen column densities and reddening parameters along the lines of sight are given in Tables \ref{hydrogentable} and \ref{reddeningtable}, respectively.

\subsection{{\it FUSE} Data}
\label{ss:FUSEdata}
Data are collected with the {\it FUSE} satellite by four channels, two with a lithium fluoride detector (LiF1 and LiF2) and two with a silicon carbide detector (SiC1 and SiC2).  Each channel contains two data segments, designated as A and B, covering adjacent wavelength regions.  The lithium fluoride channels cover the wavelength region from 989-1188 \AA{}, while the silicon carbide channels cover the region from 905-1104 \AA{}.

The data were reduced with the CALFUSE pipeline, version 2.4.0.  Data from different data segments are not coadded; rather, the data segment with the best signal-to-noise (S/N) is selected, with other data segments providing a check for consistency.

The resolution element of {\it FUSE} is $\sim15-20\kmpers$.  In fitting the non-Gaussian profiles exhibited by many of the lines in this study, we used a Gaussian point-spread function (PSF) with full-width at half-maximum (FWHM) of $15\kmpers$.  Our primary objective is to determine the equivalent widths of the various absorption lines (as opposed to deriving column density or $b$-value from individual lines); therefore, the results of our fits do not depend strongly on the exact PSF that is used.

\subsection{{\it HST} Data}
\label{ss:HSTdata}
Archived {\it HST} data taken by STIS were available for ten sightlines:  HD 24534, HD 27778, HD 37903, HD 147888, HD 185418, HD 192639, HD 206267, HD 207198, HD 207538, and HD 210839.  All of these data sets provide information on the NI triplet at 1200 \AA{} except the data for HD 24534.

We used on-the-fly calibrated STIS data \citep{Micol}.  In many cases we had multiple observations.  We coadded observations of similar echelle order, adding errors in quadrature.  We used empirically measured STIS PSF's (S. V. Penton, private communication) appropriate to the grating and aperture of the observations in the fitting of the non-Gaussian profiles that are always exhibited by the 1200 \AA{} triplet.  The PSF's we use are definitively non-Gaussian and in other contexts might be worth more discussion; however, they are narrow enough compared to the profiles that the fits and resulting equivalent widths for these very strong lines do not depend strongly on the precise nature of the PSF that is used.

For many of these sightlines, {\it IUE} data are available, covering a wavelength region similar to that of {\it HST}.  However, the few lines that are in the {\it IUE} wavelength region are blueward of Lyman-$\alpha$ and subject to considerable uncertainty that we deem unacceptable for inclusion in this study.

\section{OBSERVED NI ABSORPTION LINES}
\label{s:lines}
We follow the same procedure outlined in our previous work on OI \citep{Jensen}.  Specifically, we measure the equivalent widths of as many neutral nitrogen absorption lines as possible, then fit the measured equivalent widths to a curve of growth represented by a single velocity dispersion (i.e. $b$-value).  The equivalent widths of at least three absorption lines are required to produce a meaningful solution for the two free parameters of column density $N$ and $b$-value (measuring two absorption lines produces a unique but poorly-constrained solution; see \citeauthor{Jensen}).  In particular, to better constrain the solution and check for possible systematic errors due to the single $b$-value assumption, it is necessary to measure absorption lines with a range of $f$-values, at least some of which are on the linear portion of the curve of growth.

\begin{figure}[t!]
\begin{center}
\epsscale{1.00}
\plotone{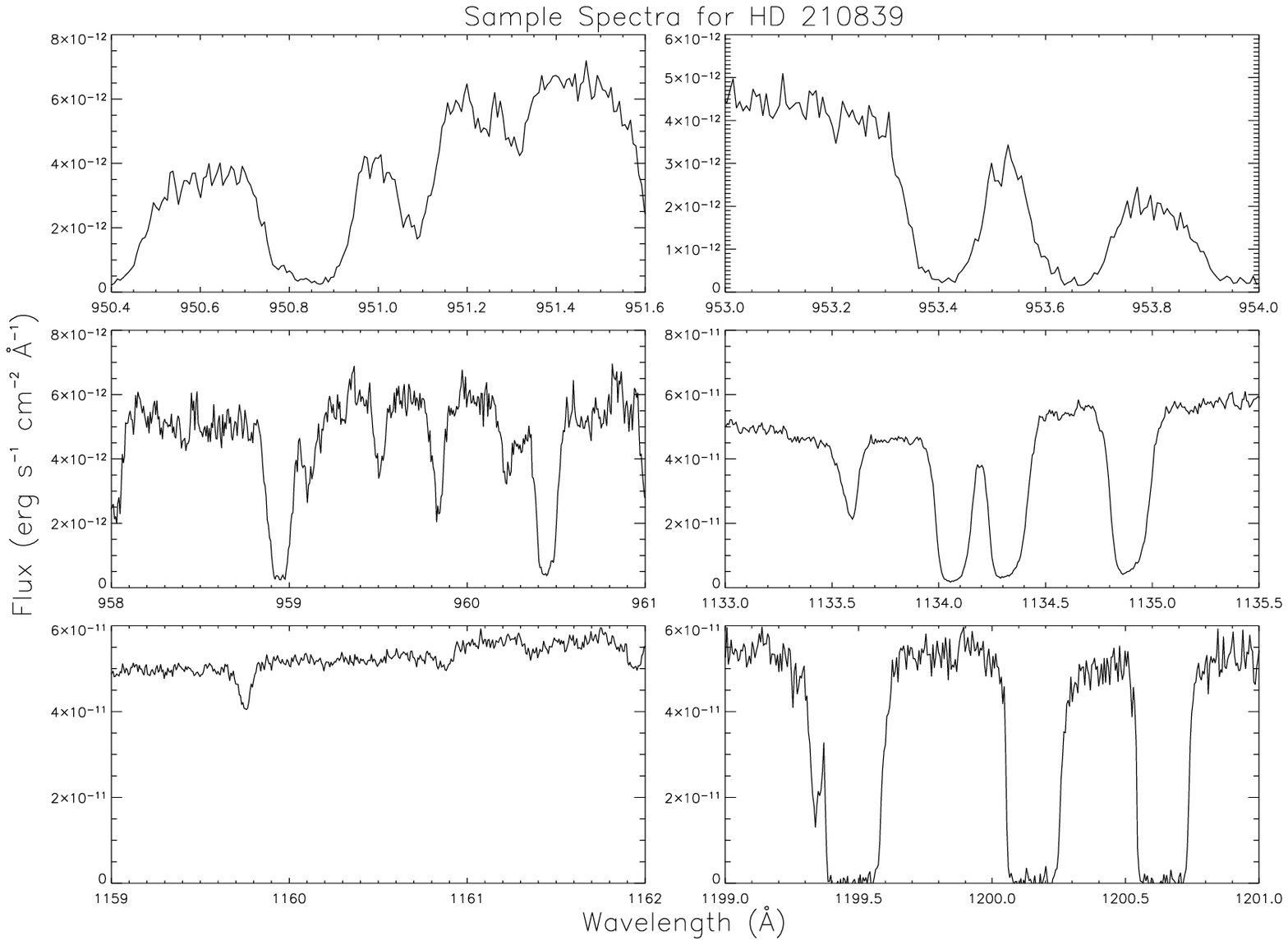}
\end{center}
\caption{The sightline toward HD 210839 has relatively high S/N, and all absorption lines used in this study are detected.  Spectra are shown unnormalized and without velocity correction.  {\bf Top Left}:---The doublet at 951 \AA{} is shown, from SiC2A {\it FUSE} data.  The two lines of the doublet are between 951.0 \AA{} and 951.5 \AA{}.  The strong feature below 951.0 \AA{} is a blend of an $\Hmol$ line and an OI line.  {\bf Top Right}:---The two visible components of a triplet at 953 \AA{} are shown, from SiC2A {\it FUSE} data.  The reddest component is destroyed by a strong $\Hmol$ feature.  {\bf Middle Left}:---The visible component of a doublet at 959 \AA{} is shown, from SiC2A {\it FUSE} data.  The line is at approximately 959.5 \AA{}.  The weaker, redder component may be seen at approximately 960.2 \AA{} but its separation from an $\Hmol$ line is unclear.  The other features are $\Hmol$ lines.  {\bf Middle Right}:---The strong triplet at 1134 \AA{} is shown, from LiF1B {\it FUSE} data.  The weaker feature at 1133.6 \AA{} is an FeII line.  {\bf Bottom Left}:---The weak doublet at 1160 \AA{} is shown, from LiF1B {\it FUSE} data.  {\bf Bottom Right}:---The strong triplet at 1200 \AA{} is shown, from {\it HST} STIS data.  An MnII line resides in the blue wing of the bluest component of the triplet.}
\label{fig:spec210839}
\end{figure}

The following is a brief description of the NI absorption lines that were used in our analysis.  Wavelengths, $f$-values, damping constants, and the instrument that was used for the observations of each of these lines are summarized in Table \ref{linetable}.  All atomic data in this section are taken from the compliation of \citet{Morton03}; the $f$-values relevant to this study are originally found in \citet{Tachiev}.  In one line of sight (HD 210839) we detected all of the absorption lines used in this study; as an example, these absorption lines are shown in Figure \ref{fig:spec210839}.

\subsection{Lines Between 950 \AA{} and 960 \AA{}}
\label{ss:950lines}
There are three multiplets between 950 \AA{} and 960 \AA{} that are observable.  One is a doublet at 951 \AA{}.  The stronger component of this doublet was detected and measured in nine lines of sight, and the weaker component was found in five of those nine lines of sight.  Another doublet is found at 959 \AA{}.  The stronger component of this doublet is found in 13 lines of sight, while the weaker component is not detected in any lines of sight.  These three lines are all relatively weak and resolved velocity structure is not apparent; therefore, the broad instrumental profile of {\it FUSE} dominates the observed profile, and Gaussian fits were appropriate in all cases (even if the intrinsic profile is slightly saturated).

In between these two doublets is a moderately strong triplet found at 953 \AA{}.  In our lines of sight, all of which possess large amounts of $\Hmol$, the reddest component of the triplet is masked by $\Hmol$ absorption.  However, the other two components are observed.  Either Gaussian or Voigt profiles were used to fit the lines, as appropriate to the observed profiles of the lines.

\subsection{1134 \AA{} Triplet}
\label{ss:1134triplet}
The triplet found at 1134 \AA{} is the most commonly observed set of absorption lines in this study, as the lines are the strongest NI transitions found in {\it FUSE} spectra and they also lie in a spectral region with relatively high S/N.  These lines were detected in every sightline for which we were able to measure any NI absorption lines.  There are several complications in analyzing these lines.  First, in some spectra there are apparent stellar features that must be fit to get an accurate representation of the continuum.  Secondly, the separation of the two weaker (and bluer) components of the triplet is so small that the lines blend in many cases.  This problem is enhanced in cases of high $b$-value and/or column density.  Thirdly, there are cases where the lines are clearly saturated but the cores do not reach zero.  This implies the presence of scattered light, contamination from telluric emission lines, or inaccuracies in the CALFUSE pipeline reduction.  There are also known instrumental features (caused by exposure to telluric emission) in this location of the LiF2A spectra that have increased over time; thus we rely on LiF1B data, except in the earliest spectra where both LiF1B and LiF2A provide good data.

We account for the non-zero cores by fitting a quadratic through the cores of the three components of the triplet, and subtracting this from the spectrum.  This is a valid method if the non-zero cores are caused by a smooth, continuous source of contamination---which may or may not be true of scattered light or a problem in the reduction pipeline.  Telluric emission would complicate matters, because while it is fairly broad ($\sim0.3$ \AA{} for each line), it would not be as uniform as our correction assumes.  However, we are able to constrain the expected level of telluric NI emission by observing other strong airglow lines (particularly the Lyman-$\beta$ and Lyman-$\gamma$ lines of HI) and using the relative strengths found in \citet{Feldman}.  These comparisons indicate that telluric NI emission is not the dominant cause of this zero-level problem.

There are some cases where the 1134 \AA{} triplet is weak enough that the while the lines are intrinsically saturated, the observed profiles are not (i.e. the intensities at the center of the lines does not reach zero and the cores are not flat).  In these cases the instrumental profile dominates the observed profile, and therefore Gaussian fits are justified and corrections cannot be made to the zero point of the spectrum.  In the majority of cases, however, flat cores are observed and Voigt profiles are used.

In the case of HD 53367, the line profiles are slightly assymetric and not well-fit by single-velocity component profiles, either Gaussian or Voigt.  We measure the equivalent widths of the 1134 \AA{} triplet in this line of sight by a simple summation relative to the assumed continuum level, rather than the equivalent width of a fit.  Errors are estimated based on the noise level in the continuum.

\subsection{1160 \AA{} Doublet}
\label{ss:1160doublet}
The weak doublet at 1160 \AA{} has been the staple of NI studies utilizing {\it Copernicus} data.  However, most {\it HST} spectra do not extend down to this wavelength.  The one spectrum in our study that barely does (HD 147888) has inferior S/N in that region compared to the {\it FUSE} spectrum.  Thus, this doublet is measured exclusively by {\it FUSE}.

As shown in Table \ref{linetable}, the $f$-values for this doublet are significantly smaller than the rest of the absorption lines that we observe---the stronger component of the doublet has an $f$-value less than half of that of the next weakest line we observe (at 951.3 \AA{}), more than three orders of magnitude smaller than the weakest component of the 1134 \AA{}, and more than four orders of magnitude smaller than the weakest component of the 1200 \AA{} triplet.  Therefore, when one or both components are detected, this doublet provides an important constraint on the overall curve of growth.  Due to the small $f$-values of the doublet, Gaussian fits are always appropriate for both components (i.e. the observed line profiles are not saturated, show no resolved velocity structure, and the instrumental profile dominates even if there is some intrinsic saturation).  In some spectra, a stellar feature interferes with the 1159.8 \AA{} component of the doublet, but in most cases the line can still be fit.

There is some level of uncertainty in the $f$-values of these two lines \citep{Morton03}.  The theoretically calculated line strengths cited in \citet{Tachiev} differ from those in \citet{Hibbert} by approximately 0.068 dex (17\%) for the stronger line and 0.057 dex (14\%) for the weaker line.  Previous literature values show an even greater range; for instance, the theoretically calculated $f$-value of the 1159.8 \AA{} line in \citet{Cowan} is three times stronger than the astrophysically-derived empirical value in \citet{Lugger}.

For simplicity and self-consistency, we uniformly assume the \citeauthor{Tachiev} $f$-values.  If our curves of growth are dominated by the weak lines, there will be a systematic difference between our derived column densities and those of \citet{Meyer} and \citet{Knauth}.  Since the \citet{Tachiev} $f$-values for this weak doublet are larger than the \citet{Hibbert} $f$-values, we expect our derived column densities to be smaller if and when there are systematic differences.  For simplicity, in this paper the values we quote from \citeauthor{Meyer} and \citeauthor{Knauth} are adjusted downward by 17\%.

\subsection{1200 \AA{} Triplet}
\label{ss:1200triplet}
In the lines of sight with {\it HST} data, we are able to measure a triplet at 1200 \AA{}.  The lines of this triplet are stronger than all of the other nitrogen lines measured in this study.  Voigt profile fits are always used to fit these lines.  There is a line of MnII in the blue wing of the bluest line of the triplet; this MnII line is relatively weak and is removed with a simple fit (it is nearly Gaussian, convolved with our empirical STIS PSF).  A stellar feature also interferes with this blue line of the triplet in some cases.  However, we were able to reconstruct the local continuum and fit the line.

\subsection{Other Lines}
\label{ss:otherlines}
There are other neutral nitrogen lines that fall in the wavelength range covered by {\it FUSE}.  However, none of these lines was clearly detected, due primarily to interference by strong $\Hmol$ absorption (see Table \ref{nolinetable}).  There are no other UV absorption lines of nitrogen in the range covered by {\it HST}.

\subsection{Errors}
\label{ss:errors}
Errors on equivalent width measurements take two forms.  When absorption lines are fitted to a Gaussian, errors are taken from standard error propagation of the curve-fitting routine and the functional form of a Gaussian.  These errors include continuum placement error added in quadrature with the errors in the fit.  When absorption lines are fitted to a Voigt profile, errors are not clear due to the lack of a functional form (the Voigt profile is fitted iteratively and involves convolution with the instrumental resolution element).  Therefore, we estimate 1-$\sigma$ errors by first taking the difference between the fit and the data (both normalized) at each point.  The uncertainty is then estimated to be the standard deviation of this difference, multiplied by the wavelength range of the fit.  This method implicitly includes continuum placement errors.  The validity of this method of error estimation was confirmed by using both methods on absorption lines for which a Gaussian was an appropriate fit.

Potential errors from the placement of the zero level of the spectrum are not formally accounted for; however, the {\it FUSE} spectra of the lines of sight in this study have very wide $\Hmol$ bands that constrain errors in the placement of the zero level over wide ranges of the {\it FUSE} bandwidth \citep{Rachford}; in most cases, the zero level is uncertain to only a few percent of the continuum level (with results in a correspondingly small uncertainty in equivalent widths due to this effect).  We have commented above (\S\ref{ss:1134triplet}) on the particular zero-level issues encountered above with the 1134 \AA{} triplet.

\section{METHODS}
\label{s:methods}

\subsection{The Curve-of-Growth Method}
\label{ss:cogmethod}
In order to derive nitrogen column densities, we follow the same procedure outlined in our previous work on OI \citep{Jensen}.  We fitted the measured equivalent widths of the various NI absorption lines to a curve of growth with a single velocity dispersion.  Several curves of growth were constructed using the damping constants of the strongest lines (the lines of the 1200 \AA{} triplet, the 1134 \AA{} triplet, and the two observable lines of the 953 \AA{} triplet), so that each absorption line was compared to the curve corresponding to its damping constant.  We found the minimum $\chi^2$ with respect to the two free parameters of column density $N$ and $b$-value, then used the extrema of the confidence ellipses to find the 1-$\sigma$ errors in $N$ and $b$-value.  The best fits for these curves of growth are reported in Table \ref{coldensities}, with 1-$\sigma$ errors.  Also reported in Table \ref{coldensities} are the abundances of nitrogen relative to hydrogen, given in parts per million (ppm).  The errors on these nitrogen abundances are taken from error propagation done in standard (i.e. non-logarithmic) notation.  In one case, HD 41117, the large errors in the NI column density combined with standard error propagation results in lower limit errors larger than the absolute value of the nitrogen abundance, and thus we have artificially capped the lower limit to be $\NH \geq 1\ppm$.

There are potential systematic errors stemming from our assumption that the velocity structure of each line of sight is reasonably approximated by a single $b$-value.  However, when measurements from absorption lines with a wide range of $f$-values are available (our measurements typically span up to $\sim4$ orders of magnitude in $f$-value), these systematic errors should be small.  We discuss this idea in more detail in \S\ref{ss:linear}.

\begin{figure}[t!]
\begin{center}
\epsscale{1.00}
\plotone{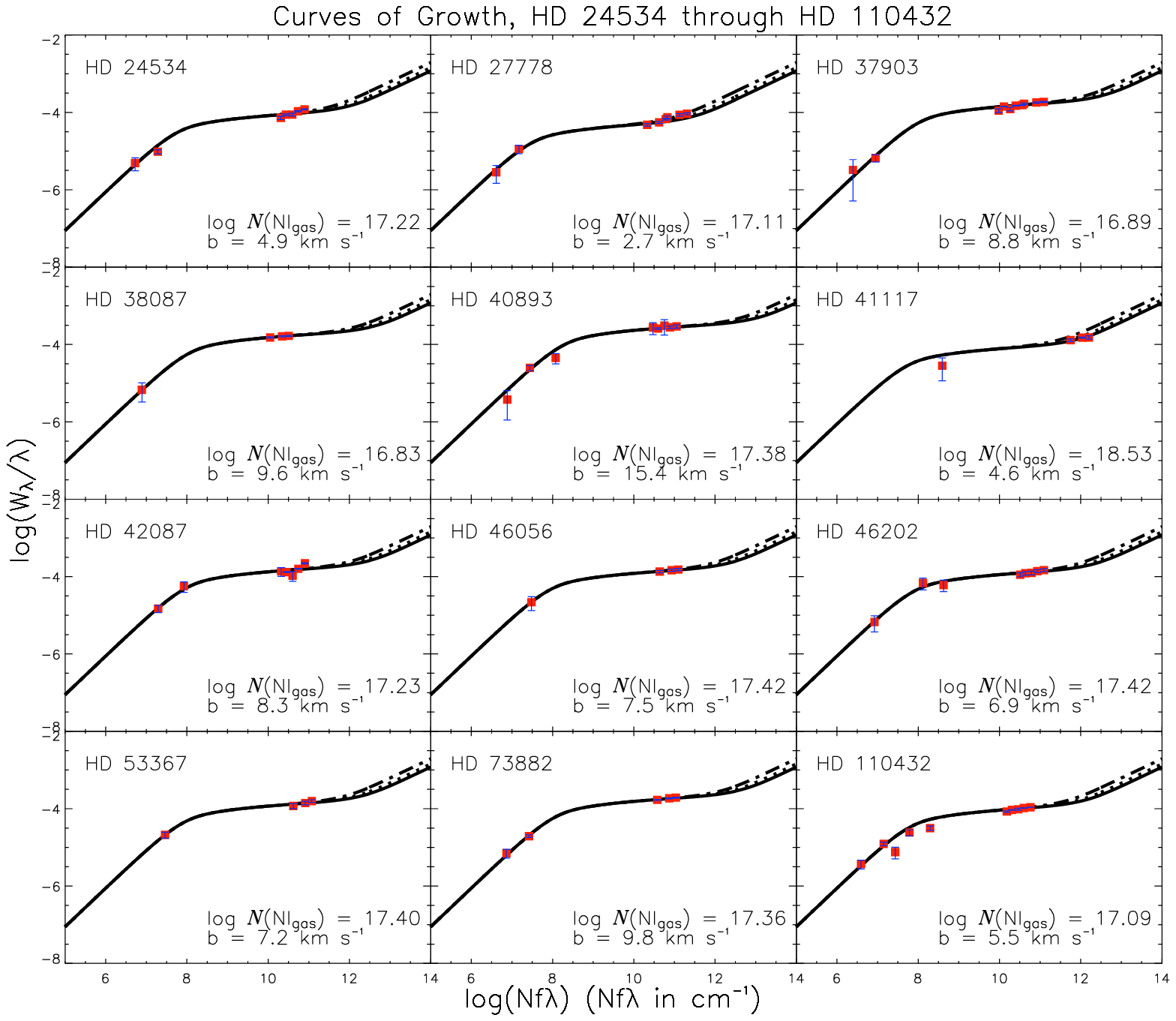}
\end{center}
\caption{Curves of growth for HD 24534 through HD 110432.  Measured equivalent widths of NI absorption lines are plotted on top of appropriate curves of growth for the adopted solution of column density and $b$-value.  Equivalent width measurements are red squares with blue error bars.  The solid, dotted, and dotted-dashed lines in the damping portion of the curve represent damping constants of $1.51\times10^8$, $2.19\times10^8$, and $4.07\times10^8$, respectively.}
\label{fig:cogs1-12}
\end{figure}
\begin{figure}[t!]
\begin{center}
\epsscale{1.00}
\plotone{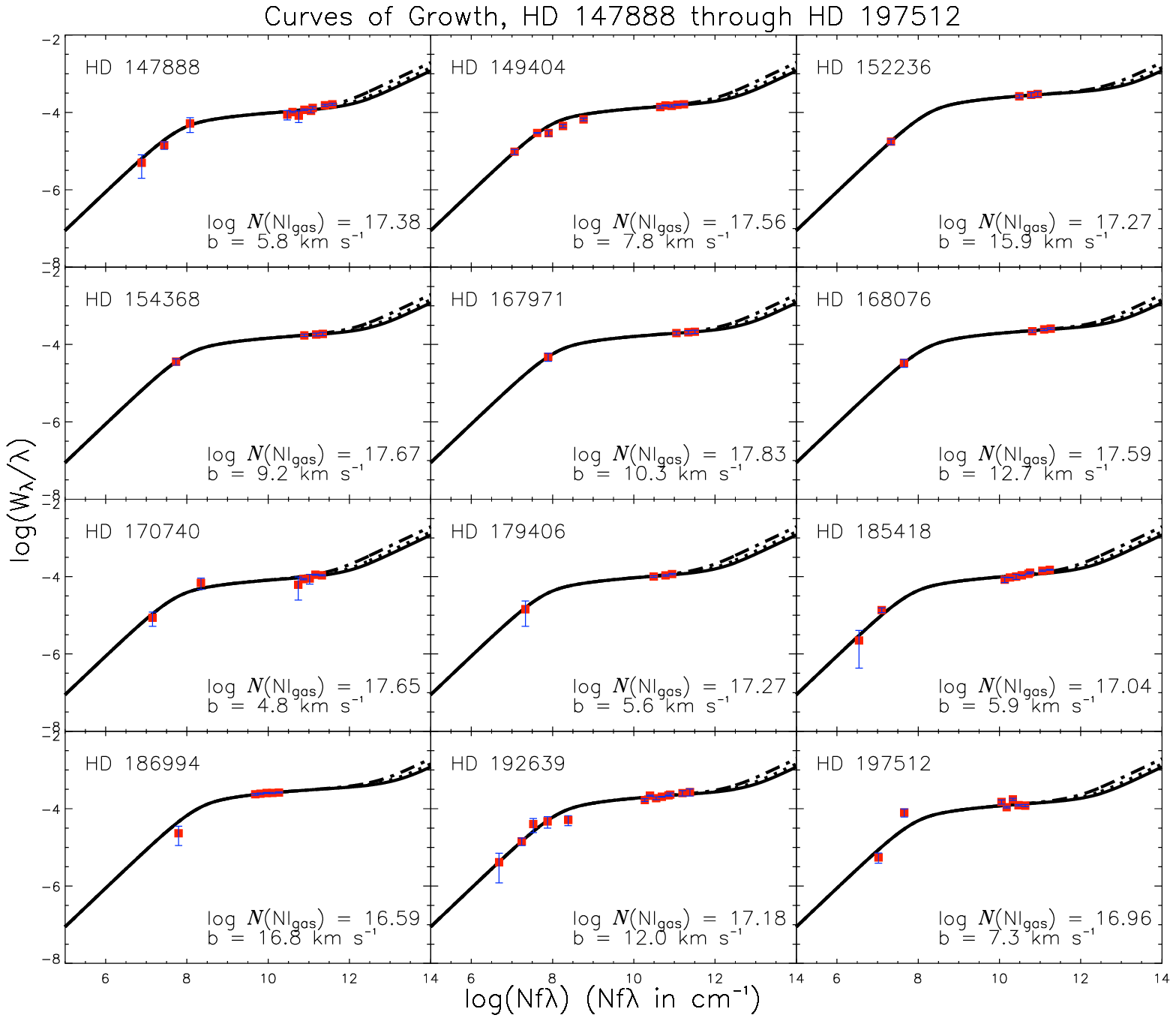}
\end{center}
\caption{Curves of growth for HD 147888 through HD 197512.  Measured equivalent widths of NI absorption lines are plotted on top of appropriate curves of growth for the adopted solution of column density and $b$-value.  Equivalent width measurements are red squares with blue error bars.  The solid, dotted, and dotted-dashed lines in the damping portion of the curve represent damping constants of $1.51\times10^8$, $2.19\times10^8$, and $4.07\times10^8$, respectively.}
\label{fig:cogs12-24}
\end{figure}
\begin{figure}[t!]
\begin{center}
\epsscale{1.00}
\plotone{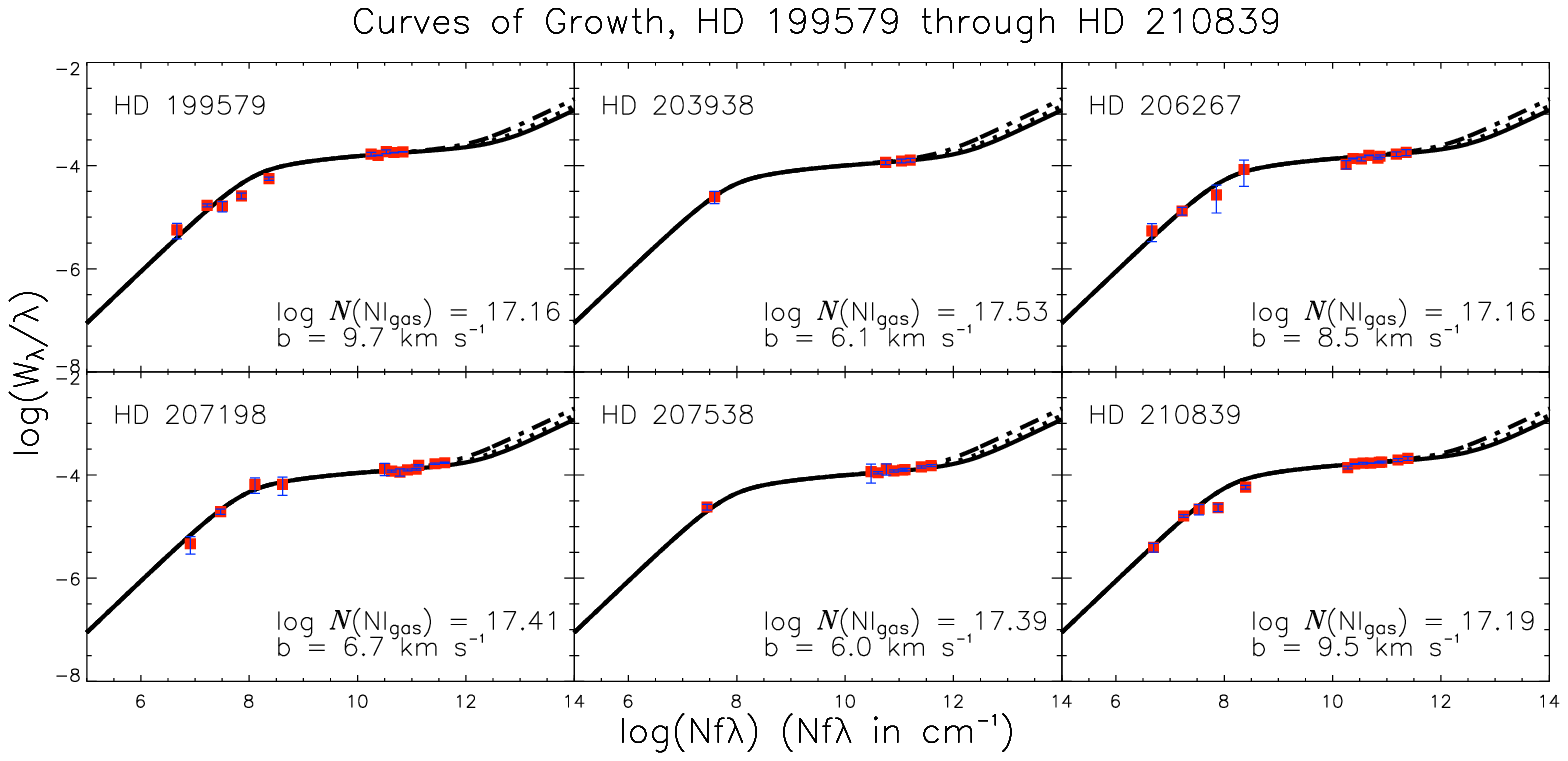}
\end{center}
\caption{Curves of growth for HD 199579 through HD 210839.  Measured equivalent widths of NI absorption lines are plotted on top of appropriate curves of growth for the adopted solution of column density and $b$-value.  Equivalent width measurements are red squares with blue error bars.  The solid, dotted, and dotted-dashed lines in the damping portion of the curve represent damping constants of $1.51\times10^8$, $2.19\times10^8$, and $4.07\times10^8$, respectively.}
\label{fig:cogs25-33}
\end{figure}

Adopted curves of growth are shown in Figures \ref{fig:cogs1-12}-\ref{fig:cogs25-33}.  In the rest of this section, we describe the measurement of hydrogen column densities and the alternative methods and consistency checks that were employed in the final determination of nitrogen column densities and $b$-values for certain lines of sight.  Again, we note here that these consistency checks rely on the assumption that the velocity structure is reasonably approximated by a single $b$-value.  However, these consistency checks are only intended to rule out very disparate results (e.g. a column density from the curve of growth that implies a damped profile in one of the strong lines when such a profile is not observed) and not to differentiate between more subtle differences in column density and velocity structure.

\subsection{Hydrogen Column Densities}
\label{ss:hydrogen}
Molecular hydrogen column densities are taken from recent {\it FUSE} studies (\citeauthor{Rachford} \citeyear{Rachford}; B. L. Rachford et al., in preparation).  In these papers $\NHmol$ is determined for each sightline by fitting several low-$J$ lines.  Atomic hydrogen densities are taken from several sources, listed in Table \ref{hydrogentable}.  We adopt literature values where they exist.  Table \ref{hydrogentable} summarizes this data.

\citet{BohlinSavDrake} showed for less reddened lines of sight that there exists a correlation between selective reddening and total hydrogen column density such that $\Htot=5.8\times10^{21}E_{B-V}$.  \citet{Rachford} showed that this correlation remains valid to approximately $\pm0.30\dex$ (a factor of 2) in $\logHtot$ even for large hydrogen column densities ($\logHtot\gtrsim21$) and large molecular fractions of hydrogen; furthermore, the correlation is much more precise than $\pm0.30\dex$ in most cases.  For two lines of sight (HD 38087 and HD 53367), we estimated total hydrogen column densities using this correlation, and then inferred atomic hydrogen column densities based on our molecular hydrogen results, that is, the atomic hydrogen column density is given by $\NHI=5.8\times10^{21}E_{B-V}-2\NHmol$.  We assume the error in $\NHI$ using this method is $\pm0.30\dex$, and the final error in the total hydrogen column density is determined from the error propagation in adding $\NHmol$ and $\NHI$; therefore lines of sight with a large molecular fraction and small error bars on $\NHmol$ will have the smallest error bars in our total estimated $\Htot$.  Similarly, for HD 179406 we quote the literature value of \citet{Hanson}, which is not a direct measurement of HI but comes from the relationship found by \citet{ShullVS}, $\NHI=5.2\times10^{21}E_{B-V}$.

$\NHI$ is determined in the remaining 27 lines of sight through profile fitting of Lyman-$\alpha$.  A few of these lines of sight warrant mentioning.  First, we made our own measurement of the Lyman-$\alpha$ profile for the line of sight toward HD 40893, using STIS data sets O8NA02010 and O8NA02020.  Our analysis of the profile of the line yields a value for the column density of $\logHI=21.50\pm0.10$.  This result is somewhat larger (by $0.22\dex$) than the result that would be obtained through use of the \citet{BohlinSavDrake} relationship, but within our estimated limits of that relationship.

Second, in lines of sight studied by \citet{Diplas}, stellar contributions to the Lyman-$\alpha$ profile were estimated.  Of the stars where we cite interstellar HI along the line of sight from \citeauthor{Diplas}, only HD 42087 is B2.5 or cooler.  \citeauthor{Diplas} deemed the estimated stellar contamination to be insignificant.

Third, we quote the values of \citet{Cartledge2} for HD 27778 and HD 147888.  These stars are cool enough that stellar Lyman-$\alpha$ might be a concern.  For HD 27778, however, the derived $\NHI$ matches, within the errors, the value estimated by \citet{Rachford} using the \citet{BohlinSavDrake} relationship.  It also results in an interstellar Kr/H abundance that precisely matches the interstellar mean.  For HD 147888, \citeauthor{Cartledge2} used HD 147933 as a proxy.  However, they report that the Lyman-$\alpha$ profile of HD 147888 results in the same column density.  This column density also agrees, within the errors, with an earlier literature value of \citet{FitzMassa}.  Thus, for HD 27778 and HD 147888 we conclude that stellar contamination is not significant.

\subsection{Alternate Methods}
\label{ss:alternate}
For each sightline, we use column density and $b$-value results from the curve-of-growth analysis to generate profiles of multiple absorption lines (in particular the 1134.98 \AA{} line and the 1160 \AA{} doublet) as a check for consistency.  These profile fits occasionally reveal flaws in the curve-of-growth analysis.  In these cases, we use other methods to determine column density.  One method we used was to assign a uniform fractional error (based on the average of the fractional errors) to each of the equivalent width measurements, then analyze the curve of growth.  This method tends to put more emphasis on the weak lines and less emphasis on the strong lines, which may be justified if there is systematic error in the strong lines that is unaccounted for.  We use this method for three sightlines (HD 24534, HD 42087, and HD 185418).  In each of these three cases, the ``uniform error'' curve-of-growth method produces a result which is more consistent with line profiles than the standard curve-of-growth method.


Occasionally, the $\chi^2$ array from our curve-of-growth analysis has multiple local minima.  In these cases, we must select which solution is most consistent with the observed profiles of the lines.  This is the case for HD 24534 and HD 53367.  In both cases we select as our solution the point which is a local minimum, but not the global minimum.  In the case of HD 24534, this is done for the ``uniform error'' curve-of-growth result.  The selected solutions and quoted errors are allowed to within 1-$\sigma$, while the other local minima are excluded on the basis of inconsistency with the observed profiles of the absorption lines.

The fact that curve-of-growth method produces two local minima indicates complex velocity structure.  For HD 24534, we noted the hints of complex velocity structure for OI \citep{Jensen}; this is even more clear in high-resolution optical KI and CaI data \citep{WeltyKI, WeltyCaI, Pan}.  We also noted above (\S\ref{ss:1134triplet}) that the profiles of lines in the 1134 \AA{} triplet were assymetric in the case of HD 53367.

\subsection{The Linear Approximation and Saturation Concerns}
\label{ss:linear}
For each line of sight we calculated the resulting column density if the weak lines of the 1160 \AA{} doublet are assumed to be on the linear portion of the curve of growth.  In one line of sight (HD 186994) we are unable to measure either component of the 1160 \AA{} doublet.  Of the remaining 29 lines of sight, 28 show consistency within the 1-$\sigma$ errors between the column density results of the curve-of-growth method and the weak-line assumption for the weakest line measured (the 1160.9 \AA{} line when measured, otherwise the 1159.8 \AA{} line).  The other line of sight, HD 197512, is consistent to 2-$\sigma$.  This demonstrates consistency in that it shows that the strong lines do not dominate the determination of column densities, and that results from using our curve-of-growth method would be similar to the results of a study if we used the weak lines only, assuming them to be unsaturated.

Systematic errors might still be a problem if the velocity structure cannot be reasonably approximated by a single-component velocity dispersion.  However, the internal consistency of most of our curves of growth is what leads us to believe that this is not the case.  When both components of the weak doublet are detected, the ratio of equivalent widths is almost always very nearly what would be from expected from assuming both lines are unsaturated ($\sim3.6$), implying little to no saturation.  For the cases where only the stronger component of the doublet is expected, the lack of a detection of the weaker line indicates that the stronger line cannot be significantly saturated.  The case where the greatest saturation (as determined from the 1160 \AA{} doublet ratio) is detected is for HD 24534, which as noted above, has a very complex velocity structure.  However, our final derived column density still places both components of the 1160 \AA{} doublet near the linear portion of the curve, and our result is consistent (within the errors) with the column density derived by \citet{Knauth} based on the weak doublet only.

To the extent that our measured sets of equivalent widths for some lines of sight do deviate somewhat from a single velocity-dispersion curve of growth (e.g. note the ``hitch'' in the data points near the linear/flat transition for the curve of growth of HD 210839), we can still be confident that the correct column density has been measured if the 1160 \AA{} doublet ratio is similar to the expectation from the unsaturated assumption.  This is true for all cases where both components of the doublet are measured except HD 24534.  In any case, determining complex velocity structure solely from {\it FUSE} data would be very imprecise at best, and we cannot do a more complex (i.e. multiple-component) curve-of-growth analysis without {\it a priori} assumptions about the velocity structure.  High-resolution optical data on KI and CaI is available for some of the lines of sight in our sample \citep{WeltyKI, WeltyCaI, Pan}, but assuming that the relative abundances of these elements to NI are constant from cloud to cloud in a given line of sight or that $b$-values are the same for all elements in a given cloud (especially given the different ionization potentials) is not more justified than assuming that the entire velocity structure is reasonably approximated by a single velocity dispersion and/or that the weak 1160 \AA{} doublet is not significantly saturated.

\section{RESULTS AND DISCUSSION}
\label{s:results}
Adopted curves of growth, with column density and $b$-value results, are shown in Figures \ref{fig:cogs1-12}-\ref{fig:cogs25-33}.  We find an error-weighted average for the N/H ratio in the ISM of $51\pm4\ppm$ (error in the mean).  This is smaller than the average N/H abundance found by \citet{Meyer} of $62^{+4}_{-3}\ppm$.  The discrepancy is more than 1-$\sigma$ but within 2-$\sigma$.  On the other hand, our results are in nearly perfect agreement to the weighted average of \citet{Knauth}, $51\pm3\ppm$.  There are a handful of lines of sight that overlap between our study and \citeauthor{Knauth} (see \S\ref{ss:compare}); the weighted average of the lines of sight that are unique to the \citeauthor{Knauth} study is $53\pm4$.  Again, both of these weighted averages have been adjusted downward by 17\% to account for the systematic difference in $f$-values.

\begin{figure}[t!]
\begin{center}
\epsscale{1.00}
\plotone{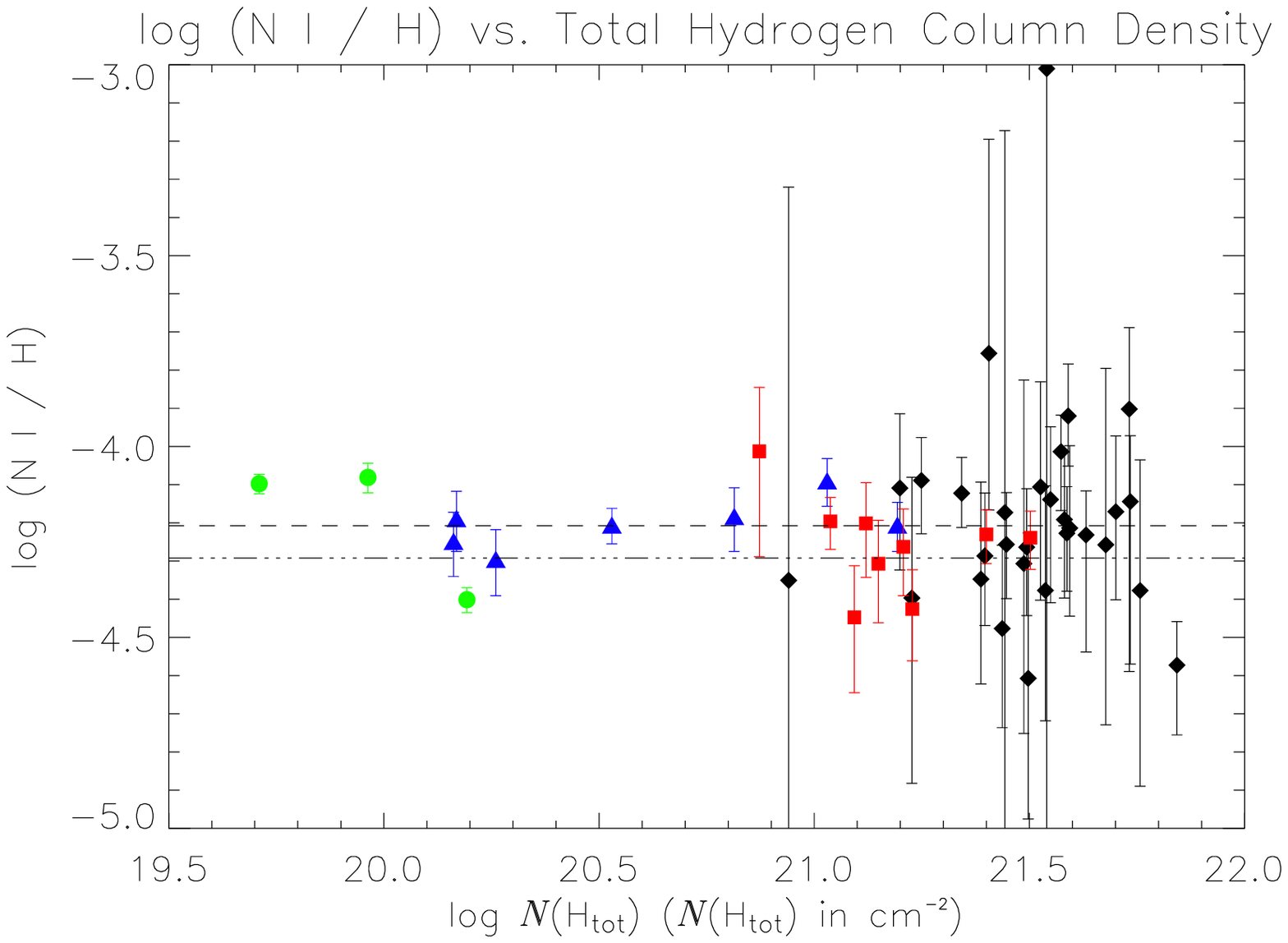}
\end{center}
\caption{A plot of $\logNH$ as a function of $\logHtot$.  {\bf Blue Triangles}:---Lines of sight from \citet{Meyer}.  {\bf Red Squares}:---Lines of sight from \citet{Knauth} not analyzed in this study.  {\bf Green Circles}:---Lines of sight from the literature \citep{Jenkins1999, Sonneborn} measured by IMAPS, included in the analysis of \citeauthor{Knauth}.  {\bf Black Diamonds}:---Lines of sight from this study.  {\bf Dashed Line}:---The mean interstellar N/H found by \citeauthor{Meyer}, $62\ppm$, is represented by the dotted line.  This is very similar to the average N/H abundance of B stars found by \citet{SnowWitt} and \citet{SofiaMeyer} discussed in \S\ref{ss:stellarNH}.  {\bf Dashed-Dotted Line}:---The average interstellar N/H of $51\ppm$ found by \citeauthor{Knauth} and this study for lines of sight with larger hydrogen column densities.  The point with the largest N/H ratio is HD 41117, which is not very well determined.  The plotted point essentially represents an upper limit.  Incorporating points from all three studies (but discarding the IMAPS points), we find a linear anticorrelation between N/H and $\Htot$ of the form $y = (-0.36\pm0.13)x + (63\pm3)$ where $y$ is N/H in parts per million (ppm) and $x$ is $\Htot / 10^{20}$.  See \S\ref{s:results} and its accompanying subsections for further discussion of these results.}
\label{fig:logNHHtot}
\end{figure}

A plot of N/H ratios as a function of total hydrogen column density is shown in Figure \ref{fig:logNHHtot}---lines of sight from \citeauthor{Knauth}, \citeauthor{Meyer}, \citet{Jenkins1999}, and \citet{Sonneborn} are included for comparison, although we do not include the \citeauthor{Jenkins1999} and \citeauthor{Sonneborn} results in our correlation or weighted average calculations.  Attempting both linear regressions and a four-parameter Boltzmann function similar to that used by \citet{Cartledge2, Cartledge3}, we do not find statistically significant correlations of the N/H ratio with regard to various extinction parameters (total extinction $A_V$, selective extinction $E_{B-V}$, or the total-to-selective ratio $R_V$) or total hydrogen volume density ($\nHvol$)---all of these calculations produce fits that are consistent with a flat line (i.e. no correlation) to within approximately 1-$\sigma$.

Two statistically significant correlations do come to light.  First, a simple linear regression of N/H vs. $\Htot$ in our data shows a negative slope that is significant to approximately 3-$\sigma$.  When we add in the lines of sight from \citeauthor{Meyer} and the unique lines of sight from \citeauthor{Knauth} (i.e. the ones that we have not independently analyzed), this anticorrelation remains, with approximately the same value of the slope and the same statistical significance.  However, by excluding HD 152236 (a line of sight with an unusual value and a small error; see \S\ref{sss:HD152236}), the significance of the correlation is significantly lessened.

We also analyzed this potential correlation with a four-parameter Boltzmann function of the form $y = A + \frac{B-A}{1+e^{\frac{x-x_0}{m}}}$, where in this case $y$ is the nitrogen abundance, $x$ is the hydrogen column density, $x_0$ is a transition value in the hydrogen column density, $m$ is the local slope of the transition, and $A$ and $B$ are the asymptotic values of $y$ for high and low column density, respectively.  Using this formula, the best fit, with a reduced $\chi^2$ of 0.138, has an abrupt transition at $\logHtot=21.05$ with values of $48\ppm$ and $60\ppm$ for the high and low column density limits, respectively.  Despite the small $\chi^2$, however, the errors in the two asymptotic values overlap.

The second correlation is with the molecular fraction of hydrogen, $\fHmol$.  We observe a positive correlation between N/H and $\fHmol$ with a slope statistically significant to approximately 3-$\sigma$ based merely on our own data.  However, this correlation disappears when we add in the \citeauthor{Meyer} and \citeauthor{Knauth} results, and can mostly likely be disregarded.

In an attempt to further investigate the potential correlation with total hydrogen column density, we analyzed previous $Copernicus$ equivalent width measurements from \citet{Bohlin83}, with the goal of exploring lower column density lines of sight \citep[these are the same equivalent widths analyzed by][]{York}.  To avoid systematic errors, we used the same $f$-values and curve-of-growth methods that we used to analyze our sample.  However, the results showed a significant amount of scatter, and there are some potential systematic errors, such as interference from stellar lines.  Additionally, some of the equivalent width measurements in certain lines of sight in the \citet{Bohlin83} data are not self-consistent, with some weak lines measured to have greater equivalent widths than other lines with stronger $f$-values.  The table in \citeauthor{Bohlin83}~that probes smaller column densities is Table 4.  However, Table 4 only utilizes a triplet of NI at 952 \AA{}.  Our analysis of the lines of sight in this table do show small abundances relative to hydrogen (a weighted average of $\sim20\ppm$ for 10 lines of sight).  However, there are potential systematic errors for two reasons.  First, there may be errors in the equivalent width measurements, as $Copernicus$ suffered from significant scattered light in the region below 1000 \AA{}.  Scattered light would systematically reduce the equivalent widths, so it is not surprising that these results are small.  Secondly, because we did not use the 952 \AA{} triplet in our more reddened lines of sight (due to interference from $\Hmol$), we do not have any examples in our data of the self-consistency (or lack thereof) of the 952 \AA{} triplet with the lines we did use.  Any systematic error between the 952 \AA{} triplet $f$-values and the $f$-values of the lines used in this paper would lead to similar systematic error between the final results for the two samples.

Many other sources in the literature are also not easily compared to our results for a two reasons.  First, individual equivalent widths, column densities, and/or abundances are not always explicitly given.  Secondly, as with Table 4 of \citet{Bohlin83}, different absorption lines are measured due to the differences in column density.  One example of this is \citet{Lugger}, who measured the nitrogen content in 10 lines of sight.  Though \citet{Lugger} used the 1134 \AA{} and 1200 \AA{} triplets, they did not measure the same weak lines as we did in this study except in a few cases.  Even the additional literature values \citep{Jenkins1999, Sonneborn} cited in \citeauthor{Knauth} do not meet this criterion, having measured lines such as the 952 \AA{} triplet (as well as the 1134 \AA{} triplet, though mostly to analyze structue).  The \citeauthor{Meyer} sample, although somewhat small, is the best sample that meets the following criteria:  (1) smaller column densities ($\logHtot \sim 20-21$) that are not too small such that ionization might be significant and (2) measurement of at least some of the same weak lines.  Lines of sight with even smaller column densities are generally shorter lines of sight, where the effects of ionization in the LISM/Local Bubble become significant.

\citeauthor{Knauth} concluded that nitrogen was depleted in lines of sight with total hydrogen column density of $\logHtot \gtrsim 21$.  This was based on a student's $t$-test; \citeauthor{Knauth} calculated the probability that the two samples (the $\logHtot < 21$ sample and the $\logHtot > 21$ sample) come from the same parent sample was 7.6\%.  This probability was lessened to 1.3\% by the exclusion of the outlying point HD 36486 ($\delta$ Ori), from \citet{Jenkins1999}.  However, the standard form of the $t$-test compares the standard deviations of the two samples but does $not$ take into account errors in the individual points.  \citeauthor{Knauth} do not explicitly state if and how they take the individual errors into account.  Similar $t$-test calculations with our sample fail to yield results similar to those of \citeauthor{Knauth}.  However, it is entirely possible this is an artifact of the broader range of errors found in our sample.  Since the $t$-test does not account for these errors, lines of sight in our sample that deviate from the mean but also have a large error are weighted more in the $t$-test than is justified.

Given that the $t$-test is not appropriate for our sample, in addition to the correlation fits discussed above, we examined the weighted averages of N/H in different regimes of hydrogen column density.  We have already discussed above the differences in our weighted average of N/H vs. that of \citeauthor{Meyer} and \citeauthor{Knauth}---the weighted averages for our results and those of \citeauthor{Knauth} differ from the weighted average of the \citeauthor{Meyer} lines of sight by more than 1-$\sigma$.  The difference in $\Htot$ between these samples is significant; 29 of our 30 lines of sight have more hydrogen than the highest column density lines of sight in \citeauthor{Meyer} (HD 143275 and HD 149757); 22 of our 30 lines of sight have a hydrogen column density that is at least a factor of two greater.  Combining the samples of \citeauthor{Meyer}, \citeauthor{Knauth}, and this paper (using our results for lines of sight common to this paper and \citeauthor{Knauth}), we compare the average N/H in lines of sight of line with $\logHtot < 21$ against those with $\logHtot > 21$; the weighted averages overlap slightly within the errors.  If we increase the breaking point to $\logHtot=21.1$, the averages cease to overlap within the errors.  At a cutoff of $\logHtot=21.4$, which roughly halves the combined sample, the weighted average is $59\pm3\ppm$ for lines of sight with $\logHtot < 21.4$, and $49\pm4\ppm$ for lines of sight with $\logHtot > 21.4$.

We have not included the \citet{Jenkins1999} and \citet{Sonneborn} results in these fits and weighted average calculations because they are measured using largely different absorption lines, as discussed above.  If we do include those results, these correlations and weighted averages generally remain unaffected; if we include the two \citet{Sonneborn} lines of sight but exclude the outlying point HD 36486 (as \citeauthor{Knauth} did to show increased statistical significance), the correlations and differences between the weighted averages are strengthened somewhat.


Based on these weighted averages, combined with the anticorrelation with $\Htot$ discussed above, we conclude that that we are seeing some evidence, however limited, that nitrogen depletion does increase in lines of sight with total hydrogen larger column densities.  This supports the conclusion of \citeauthor{Knauth}, but in both cases there are reasons for caution.  Our statistical measures are greater than 1-$\sigma$ but in some cases not by much.  Furthermore, one line of sight that influences these results is HD 152236.  As discussed in \S\ref{sss:HD152236}, this line of sight may have an unusually low N/H ratio due to nucleosynthetic effects rather than cloud conditions and chemistry (incorporation into grain material, direct depletion onto grains, and/or incorporation into nitrogen-bearing molecules).  We will proceed in this paper with the tentative conclusion that nitrogen is mildly depleted in lines of sight with greater column density---noting the uncertainties---and discuss possible reasons for this variation, as well as some side issues.


\subsection{HD 152236}
\label{sss:HD152236}
Most of the individual results we adopt are not substantially different from the previously measured abundance of nitrogen in the interstellar medium.  The primary exception is our result for HD 152236.  Our curve-of-growth analysis measures an N/H ratio of $26^{+8}_{-9}\ppm$ (several $\sigma$ below the \citeauthor{Meyer} mean, even after accounting for any systematic differences in $f$-values) and a $b$-value of $15.9\pm1.4\kmpers$.  While this is an unusually large $b$-value, it is consistent with what we previously found for OI in this line of sight \citep{Jensen}, as well as the line profiles of both NI and OI.  What is unusual, however, is that the measured column density for OI in our previous paper was very large, leading to the largest overall O/H ratio in that sample, while here the measured N/H ratio is the second smallest in our sample (the line of sight toward HD 37903 has a smaller N/H ratio but with a much larger upper limit).  The consistency of the $b$-values leads us to believe that even if there is systematic error in the N/H or O/H ratios, there is unlikely to be systematic error in NI column density relative to OI or vice versa, and the N/O ratio must be very small.  Using the adopted numbers in this paper and \citet{Jensen}, N/O = $1.2\times10^{-2}$, compared to the solar  \citep{Holweger} and ISM \citep{Knauth} ratios where ${\rm N/O}\sim0.2$.  Even if a smaller column density for OI is accepted (as discussed in \citeauthor{Jensen}), N/O is unlikely to be any larger than 0.05.

It is unclear what process or processes in the ISM could cause such a low N/O ratio.  To explain stellar abundances, \citet{Walborn} suggested that some members of the Sco OB1 association, of which HD 152236 is a member, may have formed from nitrogen-deficient material; this result may be evidence for that suggestion.  Despite this piece of corroborating evidence, the ISM processes or chemistry that may have initially created this enhanced depletion are not immediately obvious.  Furthermore, it is not known to what extent the abundances of an OB star trace the abundances of the intervening ISM.  Whether HD 152336, with the largest total hydrogen column density in our sample, shows such a low N/H ratio due to nucleosynthetic effects or cloud conditions and chemistry is ultimately unclear.

\subsection{Comparison With Previous Work}
\label{ss:compare}
Our sample overlaps with the sample of \citet{Knauth} for eight lines of sight:  HD 24534, HD 73882, HD 110432, HD 147888, HD 179406, HD 185418, HD 192639, HD 210839.  Our final adopted N/H ratios agree with those of \citeauthor{Knauth} for all eight sightlines within the errors, even after potential systematic errors from differing $f$-values are considered.  It is worth noting that our method of adopting of column densities for HD 24534 and HD 185418 was different than for the other sightlines in this sample (see \S\ref{ss:alternate}).  In general, our error bars are larger than those found by \citeauthor{Knauth}; this is likely an artifact of our use of the curve-of-growth method as opposed to the determination of column densities from the weak doublet only.

\subsection{$b$-values}
\label{ss:bvalues}
It has long been assumed \citep{York} that interstellar OI and NI are coincident in position and velocity space.  If this is true, the two elements should have the same $b$-value, and if a $b$-value can be measured for one element that same $b$-value can be assumed for the other element.  Using the same method and data sets as our previous work on OI \citep{Jensen} allows us to compare our results for the independently measured $b$-values of OI and NI.  Of the 24 common sightlines for which we report both OI and NI column densities (we did not report OI results for HD 40893, HD 41117, HD 147888, HD 149404, HD 167791, and HD 203938), we find 1-$\sigma$ agreement for $b$-values for 15 lines of sight.  Another six lines of sight (HD 24534, HD 37903, HD 38087, HD 154368, HD 170740, and HD 206267) fall outside the region of 1-$\sigma$ agreement but agree within 2-$\sigma$.  Three lines of sight (HD 38087, HD 199579, and HD 207198) only possess agreement within 3-$\sigma$.  Therefore, with the possible exception of a few cases, disagreement in $b$-values is likely statistical.

The three lines of sight with only 3-$\sigma$ agreement are the most likely to be cases of either systematic errors somewhere in the measurement of absorption lines (affecting the curve of growth and subsequent calculation of the $b$-value) or real cases where OI and NI are not in the same clouds and/or do not have the same physical conditions.  In the case of HD 38087, the column density and $b$-value for OI we derived in \citet{Jensen} are very uncertain, based on only two absorption lines.  It is probable that these values are more uncertain than the quoted errors.  Therefore, the discrepancy in $b$-values for NI and OI for HD 38087 cannot be considered significant.

HD 199579 was flagged as a potential outlier in \citet{Jensen}.  A very large $b$-value and a very small oxygen column density (leading to a small O/H ratio) were measured.  However, profile fitting of the 1039 \AA{} absorption line of OI seemed to favor a larger column density and smaller $b$-value of $\sim10\kmpers$---consistent with our $b$-value measured for nitrogen.  Since our measurement of NI for HD 199579 does favor this $b$-value and does have weak lines constraining the linear portion of the curve of growth, it is probably best to adopt the NI $b$-value.

The situation is somewhat more complex for HD 207198.  The NI curve of growth is relatively straightforward, with several strong lines constraining the flat portion of the curve, and several weak lines constraining the linear portion and the transition region between linear and flat.  The OI curve of growth is somewhat less certain visually, though the statistical analysis constricts the range of possible $b$-values.  In \citet{Pan}, optical data on KI shows multiple cloud components.  However, the cloud components are roughly symmetric and heavily blended.  The total KI profile (i.e. approximating the blended profiles of multiple cloud components as one profile) has a FWHM of $\sim12\kmpers$, which would correspond to a $b$-value of $\sim7\kmpers$.  If a $b$-value is determined empirically for each absorption line of OI in the HD 207198 spectrum, these empirical $b$-values increase with line strength.  This indicates a complex line of sight with small clouds that become important in the stronger lines.  It would be difficult to perform meaningful analysis of this line of sight using a multiple-component profile, due to the limits of the resolution of {\it FUSE} ($\sim15-20\kmpers$).  However, the $b$-value of all cloud components taken as a whole is probably better represented by the $\sim7\kmpers$ $b$-value that we estimate from \citet{Pan} and our NI results than our previous results for OI \citep{Jensen}.


\subsection{N/H of Galactic Stars}
\label{ss:stellarNH}
The two current solar N/H ratios that are widely accepted are $93\pm16\ppm$, from \citet{Grevesse}, and the more recent work of \citet{Holweger}, who found (N/H)$_{\odot}=85\pm22\ppm$.  As with many other elements in the ISM, however, the solar value may not be an appropriate abundance standard.  \citet{Lodders}, for example, suggests a pre-solar nebula N/H abundance of $79^{+23}_{-18}\ppm$.  N/H abundances found in B stars further complicate the situation:  \citet{SnowWitt} find average stellar N/H ratios of $63\ppm$ (field B stars) and $68\ppm$ (cluster B stars), while \citet{SofiaMeyer} find an average N/H of $65\ppm$.  As discussed in \citet{Meyer}, it is therefore likely that B star abundances are a better standard for N/H in the ISM than the solar abundance.  These surveys did not measure N/H in the same samples of F and G stars where other abundances were measured.  Therefore, the N/H abundance of B stars would seem represent a better potential ISM standard than the solar N/H abundance; however, we do not have the full picture, as it is unclear how the N/H abundance of F and G stars measures up.  It is also worth noting that B stars seem to be a worse potential abundance standard than F and G stars for many other elements \citep{SofiaMeyer}.

If the N/H abundances of B stars are adopted as the ISM standard, the situation is rectified somewhat:  our results would indicate that $14\ppm$ of nitrogen is missing from the gas-phase ISM, without taking into account the errors.  The results of \citet{Knauth} would indicate a similar gas-phase nitrogen deficiency.  On the other hand, the lines of sight in \citeauthor{Meyer} appear to have very little gas-phase NI depletion.  However, the errors in B star abundances as determined by \citet{SofiaMeyer} are quite large ($34\ppm$).  Additionally, these inferred depletions change with differing $f$-values for the weak lines, as we have noted many times already (the derived stellar N/H abundances are not dependent on these $f$-values).  It is therefore difficult to draw any firm conclusions regarding the amount of nitrogen in dust, even if a B star abundance is adopted.  Even if the absolute amount is not known, however, the difference in N/H of $\sim10\ppm$ for lines of sight with larger hydrogen column densities that our results hint at must be explained.  We briefly discuss a few possibilities in \S\ref{ss:dust}.

As discussed in many previous papers \citep[e.g.,][]{SnowJGR, SofiaMeyer, Jensen}, it may be the case that good abundance standards in the ISM do not exist.  Processes that occur during the formation of stars such as ambipolar diffusion, photoevaporation of the disk, and late metallicity pollution from the infall of protoplanets are likely to have a great impact on the details of each individual star, explaining why the variation in stellar abundances is greater than the variation of the well-mixed ISM from which they formed.  If a better understanding of an appropriate ISM standard for N/H can be gained, however, we may be able to better understand just where the NI goes in the densest sightlines.

\subsection{Ionization Effects}
\label{ss:ionization}
It is fair to ask whether or not any of our results, or the results of previous studies, may be influenced by ionization effects.  Indeed, \citet{Moos} concluded that ionization is responsible for a deficiency of NI relative to HI in the local interstellar.  However, \citet{Jenkins} discuss the processes that couple the ionization fractions of nitrogen and oxygen to hydrogen.  With a larger ionization cross-section, it might be expected that nitrogen is more ionized than hydrogen.  However, the ionization fractions of nitrogen and hydrogen are also governed by a charge-exchange reaction \citep{ButlerDalgarno}.  Only if $\nelec \gg \nHvol$ will nitrogen and hydrogen show a significant departure from the very similar ionizations assumed in \S\ref{s:intro}.  Figure 2 of \citet{Jenkins} shows a model that indicates that nitrogen will be substantially more ionized than hydrogen only if $\logHI \ll 19$.  However, in lines of sight toward four white dwarfs \citeauthor{Jenkins} did find a smaller abundance of NI than expected.  This indicates potential problems with the model, possibly because high-energy photons from helium recombination are a large influence (the model assumes only photoionization from hot stars and a hot-gas conduction front).

If there is a potential problem with the \citet{Jenkins} model, it is not clear how this should affect the assumption that ionization is not important in our lines of sight.  The lines of sight in this study have hydrogen column densities at least two orders of magnitude larger than the four lines of sight in the \citeauthor{Jenkins} study; the column densities in the \citet{Meyer} lines of sight are at least one order magnitude larger.  The reddening of our lines of sight also imply fairly large dust content, which might contribute to shielding the gas clouds from higher-energy photons.  Our lines of sight are also not necessarily subject to the unique conditions of the LISM/Local Bubble.

An additional consideration is that all of the lines of sight in this study have $\fHmol \geq 0.09$, including nearly half with $\fHmol \geq 0.4$.  Lines of sight with such large molecular fractions of hydrogen are typically indicative of environments that are largely self-shielded.  In seven lines of sight spanning several orders of magnitude in $\fHmol$, \citeauthor{Meyer} found no evidence of increasing depletion of nitrogen as the molecular fraction of hydrogen increases \citep[see Figure 2 of][]{Meyer}.  In fact, the same figure shows very limited evidence that nitrogen is more depleted in lines of sight with smaller $\fHmol$---this would be consistent with the possibility that ionization effects are important.  N/H follows a similar pattern with respect to total hydrogen column density as well---the two lines of sight with the smallest N/H also have two of the three smallest hydrogen column densities.  However, it is worth noting that the \citeauthor{Meyer} sample, at a total of seven lines of sight, is very small.

\subsection{Dust Grains and Sinks of NI}
\label{ss:dust}
Given that the results of this paper and \citet{Knauth} point (however tentatively) toward an increase in nitrogen depletion in dense lines of sight, the question must be asked:  where does the NI go?  \citeauthor{Knauth} determined that $\Nmol$ could not be a significant sink for NI in the lines of sight in that study.  Furthermore, unlike oxygen, silicon, and iron, nitrogen is not a significant ingredient in most dust models---a moderate amount of nitrogen is required for models that include organic refractory material \citep{SnowWitt}, but nitrogen is not necessary for dust models dominated by silicates and oxides.  Perhaps, as \citeauthor{Knauth} suggest, our understanding of interstellar nitrogen chemistry is incomplete.  Another possibility is that our observations are probing lines of sight with one or more regions where the local density is high enough to see a fair amount of atomic nitrogen depleted directly onto dust grains, as opposed to being an ingredient in the molecules that make up grains.  In this case no further understanding of nitrogen chemistry is needed.

\section{SUMMARY}
\label{s:summary}
We measured the NI column densities and $b$-values of 30 lines of sight with $\logHtot \gtrsim 21$ using measurements of several NI absorption lines and fitting those to curves of growth.  We find a weighted-average value for N/H in these sightlines of $51\pm4\ppm$.  Within the uncertainties, this is similar to the $f$-value adjusted ratio of $51\pm3$ found by \citet{Knauth}, but smaller than the $f$-value adjusted ratio of $62^{+4}_{-3}$ found by \citeauthor{Meyer} using less reddened lines of sight.  We also note a trend with respect to $\Htot$ when all three samples are considered; similar statistically significant trends do not exist for extinction parameters or other parameters such as $\fHmol$ and $\nHvol$.  We take these results to be limited evidence that the conclusions of \citeauthor{Knauth} are correct, that the gas-phase depletion of nitrogen relative to hydrogen is enhanced in lines of sight where $\logHtot \gtrsim 21$.  Potential sinks of NI in the ISM are unclear.  The reason for nitrogen depletion may be the direct accumulation of nitrogen onto large grains---an indication that perhaps the lines of sight observed are nearing the physical characteristics of the ever-elusive ``translucent'' clouds.  Continued improvement in observational capability (such as with the $Cosmic$ $Origins$ $Spectrograph$) may allow these clouds to finally be detected.  Finally, instead of explanations that invoke dust models or nitrogen chemistry, we cannot rule out nucleosynthetic effects as playing a significant role in these variations.  Our prime example of this possibility is HD 152236, where low N/H and N/O ratios in the line of sight may be correlated (as either a cause or an effect) with similar ratios in the Sco OB1 association.

\acknowledgments
This work has been supported by NASA contracts NAG5-12279 and NAS5-32985.  We would like to thank the referee for many insightful comments and suggestions.  We would also like to thank C. W. Danforth and B. A. Keeney for many helpful discussions and IDL programs that we have adapted to our purposes, and S. V. Penton for the use of his empirically measured STIS PSF's.


\clearpage \clearpage


\clearpage \clearpage

\clearpage \clearpage

\clearpage \clearpage

\clearpage \clearpage

\clearpage \clearpage


\begin{deluxetable}{cccccc}
\tablecolumns{6}
\tablewidth{0pc}
\tabletypesize{\footnotesize}
\tablecaption{Sightlines: Stellar Data\label{stellardata}}
\tablehead{\colhead{Star Name} & \colhead{Distance (pc)} & \colhead{Distance Reference\tablenotemark{a}} & \colhead{l} & \colhead{b} & \colhead{Spectral Type}}
\startdata
HD 24534 & 590 & 2 & 163.08 & -17.14 & O9.5pe \\
HD 27778 & 223 & 1 & 172.76 & -17.39 & B3V \\
HD 37903 & 910 & 2 & 206.85 & -16.54 & B1.5V \\
HD 38087 & 480 & 2 & 207.07 & -16.26 & B5V \\
HD 40893 & 2800 & 2 & 180.09 & +4.34 & B0IV \\
HD 41117 & 1000 & 2 & 189.69 & -0.86 & B2Iae \\
HD 42087 & 1200 & 2 & 187.75 & +1.77 & B2.5Ibe\\
HD 46056 & 2300 & 2 & 206.34 & -2.25 & O8V \\
HD 46202 & 2000 & 2 & 203.31 & -2.00 & O9V \\
HD 53367 & 780 & 2 & 223.71 & -1.90 & B0IVe \\
HD 73882 & 1100 & 2 & 260.18 & +0.64 & O8V \\
HD 110432 & 301 & 1 & 301.96 & -0.20 & B2pe \\
HD 147888 & 136 & 1 & 353.65 & 17.71 & B5V \\
HD 149404 & 820 & 2 & 340.54 & +3.01 & O9Iae \\
HD 152236 & 1800 & 3 & 343.03 & +0.87 & B1Ia+pe \\
HD 154368 & 960 & 2 & 349.97 & +3.22 & O9Ia \\
HD 167971 & 730 & 2 & 18.25 & +1.68 & O8e \\
HD 168076 & 2100 & 2 & 16.94 & +0.84 & O5f \\
HD 170740 & 213 & 1 & 21.06 & -0.53 & B2V \\
HD 179406 & 160 & 2 & 28.23 & -8.31 & B3V \\
HD 185418 & 950 & 2 & 53.6 & -2.17 & B0.5V \\
HD 186994 & 2500 & 2 & 78.62 & +10.06 & B0III \\
HD 192639 & 1100 & 2 & 74.90 & +1.48 & O8e \\
HD 197512 & 1700 & 2 & 87.89 & +4.63 & B1V \\
HD 199579 & 1200 & 2 & 87.50 & -0.30 & O6Ve\\
HD 203938 & 700 & 2 & 90.56 & -2.23 & B0.5IV \\
HD 206267 & 1000 & 3 & 99.29 & +3.74 & O6f \\
HD 207198 & 1000 & 2 & 103.14 & +6.99 & O9IIIe \\
HD 207538 & 880 & 2 & 101.60 & +4.67 & B0V \\
HD 210839 & 505 & 1 & 103.83 & +2.64 & O6If \\
\enddata
\tablenotetext{a}{References:---(1) Hipparcos parallax of 4-$\sigma$ precision or better.  (2) Spectroscopic distance modulus.  (3) Member of an OB association, cluster, or multiple-star system.}
\end{deluxetable}

\clearpage \clearpage

\begin{deluxetable}{ccccccccc}
\tablecolumns{9}
\tablewidth{0pc}
\tabletypesize{\scriptsize}
\tablecaption{Sightlines: Hydrogen Data\tablenotemark{a}\label{hydrogentable}}
\tablehead{\colhead{Sightline} & \colhead{$\logHmol$} & \colhead{Ref.\tablenotemark{b}} & \colhead{$\logHI$} & \colhead{HI Method\tablenotemark{c}} & \colhead{Ref.\tablenotemark{d}} & \colhead{$\logHtot$} & \colhead{$\nHvol$} & \colhead{$\fHmol$}}
\startdata
HD 24534 & $20.92\pm0.04$ & 1 & $20.73\pm0.06$ & Ly-$\alpha$ & 1 & $21.34\pm0.03$ & 1.21 & 0.76 \\
HD 27778\tablenotemark{e} & $20.79\pm0.06$ & 1 & $21.10\pm0.10$ & Ly-$\alpha$ & 2 & $21.40\pm0.06$ & 3.62 & 0.49 \\
HD 37903 & $20.92\pm0.06$ & 2 & $21.17\pm0.10$ & Ly-$\alpha$ & 1 & $21.50^{+0.06}_{-0.05}$ & 1.12 & 0.53 \\
HD 38087 & $20.64\pm0.07$ & 2 & $20.91\pm0.30$ & Bohlin & 3 & $21.23^{+0.17}_{-0.13}$ & 1.14 & 0.52 \\
HD 40893 & $20.58\pm0.05$ & 2 & $21.50\pm0.10$ & Ly-$\alpha$ & 3 & $21.59^{+0.08}_{-0.09}$ & 0.49 & 0.19 \\
HD 41117 & $20.68\pm0.10$ & 2 & $21.40\pm0.15$ & Ly-$\alpha$  & 1 & $21.54^{+0.12}_{-0.11}$ & 1.12 & 0.28 \\
HD 42087\tablenotemark{e} & $20.52\pm0.12$ & 2 & $21.39\pm0.11$ & Ly-$\alpha$  & 1 & $21.49\pm0.09$ & 0.84 & 0.21 \\
HD 46056 & $20.68\pm0.06$ & 2 & $21.38\pm0.14$ & Ly-$\alpha$ & 1 & $21.53^{+0.11}_{-0.10}$ & 0.47 & 0.29 \\
HD 46202 & $20.68\pm0.07$ & 2 & $21.58\pm0.15$ & Ly-$\alpha$ & 1 & $21.68\pm0.12$ & 0.77 & 0.20 \\
HD 53367 & $21.04\pm0.05$ & 2 & $21.32\pm0.30$ & Bohlin & 3 & $21.63^{+0.17}_{-0.12}$ & 1.78 & 0.51 \\
HD 73882 & $21.11\pm0.08$ & 1 & $21.11\pm0.15$ & Ly-$\alpha$ & 4 & $21.59^{+0.08}_{-0.07}$ & 1.14 & 0.67 \\
HD 110432 & $20.64\pm0.04$ & 1 & $20.85\pm0.15$ & Ly-$\alpha$ & 5 & $21.20^{+0.08}_{-0.07}$ & 1.70 & 0.55 \\
HD 147888\tablenotemark{e} & $20.47\pm0.05$ & 2 & $21.71\pm0.09$ & Ly-$\alpha$ & 2 & $21.76\pm0.08$ & 13.63 & 0.10 \\
HD 149404 & $20.79\pm0.04$ & 2 & $21.40\pm0.14$ & Ly-$\alpha$ & 1 & $21.57^{+0.11}_{-0.10}$ & 1.48 & 0.33 \\
HD 152236 & $20.73\pm0.12$ & 2 & $21.77\pm0.13$ & Ly-$\alpha$ & 1 & $21.84\pm0.11$ & 1.25 & 0.15 \\
HD 154368 & $21.16\pm0.07$ & 1 & $21.00\pm0.05$ & Ly-$\alpha$ & 6 & $21.59\pm0.05$ & 1.31 & 0.74 \\
HD 167971 & $20.85\pm0.12$ & 1 & $21.60\pm0.30$ & Ly-$\alpha$ & 7 & $21.73^{+0.24}_{-0.20}$ & 2.40 & 0.26 \\
HD 168076 & $20.68\pm0.08$ & 1 & $21.65\pm0.23$ & Ly-$\alpha$ & 1 & $21.73^{+0.20}_{-0.18}$ & 0.84 & 0.18 \\
HD 170740 & $20.86\pm0.08$ & 1 & $21.04\pm0.15$ & Ly-$\alpha$ & 1 & $21.41^{+0.08}_{-0.07}$ & 3.87 & 0.57 \\
HD 179406 & $20.73\pm0.07$ & 2 & $21.23\pm0.15$ & Shull & 8 & $21.44^{+0.10}_{-0.09}$ & 5.61 & 0.39 \\
HD 185418 & $20.76\pm0.05$ & 1 & $21.11\pm0.15$ & Ly-$\alpha$ & 4 & $21.39^{+0.09}_{-0.08}$ & 0.83 & 0.47 \\
HD 186994 & $19.59\pm0.04$ & 2 & $20.90\pm0.15$ & Ly-$\alpha$ & 9 & $20.94^{+0.14}_{-0.13}$ & 0.11 & 0.09 \\
HD 192639 & $20.69\pm0.05$ & 1 & $21.32\pm0.12$ & Ly-$\alpha$ & 1 & $21.49^{+0.09}_{-0.08}$ & 0.90 & 0.32 \\
HD 197512 & $20.66\pm0.05$ & 1 & $21.26\pm0.15$ & Ly-$\alpha$ & 4 & $21.44^{+0.11}_{-0.10}$ & 0.52 & 0.33 \\
HD 199579 & $20.53\pm0.04$ & 1 & $21.04\pm0.11$ & Ly-$\alpha$ & 1 & $21.25\pm0.07$ & 0.48 & 0.38 \\
HD 203938 & $21.00\pm0.06$ & 1 & $21.48\pm0.15$ & Ly-$\alpha$ & 4 & $21.70^{+0.10}_{-0.09}$ & 2.32 & 0.40 \\
HD 206267 & $20.86\pm0.04$ & 1 & $21.30\pm0.15$ & Ly-$\alpha$ & 7 & $21.54^{+0.09}_{-0.08}$ & 1.12 & 0.42 \\
HD 207198 & $20.83\pm0.04$ & 1 & $21.34\pm0.17$ & Ly-$\alpha$ & 1 & $21.55^{+0.11}_{-0.10}$ & 1.15 & 0.38 \\
HD 207538 & $20.91\pm0.06$ & 1 & $21.34\pm0.12$ & Ly-$\alpha$ & 1 & $21.58^{+0.08}_{-0.07}$ & 1.40 & 0.43 \\
HD 210839 & $20.84\pm0.04$ & 1 & $21.15\pm0.10$ & Ly-$\alpha$ & 1 & $21.45^{+0.06}_{-0.05}$ & 1.79 & 0.49 \\
\enddata
\tablenotetext{a}{Same as Table 2 of \citet{Jensen}, with inclusion of new lines of sight, minor revisions, and correction of typographical errors.}
\tablenotetext{b}{References for $\Hmol$:  (1) \citet{Rachford}.  (2) B. L. Rachford et al., in preparation.}
\tablenotetext{c}{Methods for HI:  Ly-$\alpha$---Profile fitting of the Lyman-$\alpha$ line.  Bohlin---$\NHI=5.8\times10^{21}E_{B-V}-2\NHmol$ from relationship in \citet{BohlinSavDrake} and further confirmed in \citet{Rachford}; errors in $\NHI$ assumed to be $\pm0.30$ dex.  Shull---$\NHI=5.2\times10^{21}E_{B-V}$ from \citet{ShullVS}.}
\tablenotetext{d}{References where HI measurements/estimations appear:  (1) \citet{Diplas}.  (2) \citet{Cartledge2}.  (3) This paper.  (4) \citet{FitzMassa}; errors in $\NHI$ assumed to be $\pm0.15$ dex.  (5) \citet{Rachford110432}.  (6) \citet{Snow154368}.  (7) \citet{Rachford}.  (8) \citet{Hanson}.  (9) \citet{BohlinSavDrake}.}
\tablenotetext{e}{See \S\ref{ss:hydrogen} for further discussion on these stars.}
\end{deluxetable}

\clearpage \clearpage

\begin{deluxetable}{cccc}
\tablecolumns{4}
\tablewidth{0pc}
\tabletypesize{\footnotesize}
\tablecaption{Sightlines: Reddening Data\label{reddeningtable}}
\tablehead{\colhead{Sightline} & \colhead{$E_{B-V}$ (magnitudes)} & \colhead{$A_V$ (magnitudes)} & \colhead{$R_V$}}
\startdata
HD 24534 & 0.59 & 2.05 & 3.47 \\
HD 27778 & 0.37 & 1.01 & 2.72 \\
HD 37903 & 0.35 & 1.28 & 3.67 \\
HD 38087 & 0.29 & 1.61 & 5.57 \\
HD 40893 & 0.46 & 1.13 & 2.46 \\
HD 41117 & 0.45 & 1.23 & 2.74 \\
HD 42087 & 0.36 & 1.10 & 3.06 \\
HD 46056 & 0.50 & 1.30 & 2.60 \\
HD 46202 & 0.49 & 1.39 & 2.83 \\
HD 53367 & 0.74 & 1.76 & 2.38 \\
HD 73882 & 0.70 & 2.36 & 3.37 \\
HD 110432 & 0.51 & 2.02 & 3.95 \\
HD 147888 & 0.47 & 1.91 & 4.06 \\
HD 149404 & 0.68 & 2.23 & 3.28 \\
HD 152236 & 0.68 & 2.24 & 3.29 \\
HD 154368 & 0.78 & 2.34 & 3.00 \\
HD 167971 & 1.08 & 3.42 & 3.17 \\
HD 168076 & 0.78 & 2.77 & 3.55 \\
HD 170740 & 0.48 & 1.30 & 2.71 \\
HD 179406 & 0.33 & 0.94 & 2.86 \\
HD 185418 & 0.50 & 1.16 & 2.32 \\
HD 186994 & 0.17 & 0.56 & 3.29 \\
HD 192639 & 0.66 & 1.87 & 2.84 \\
HD 197512 & 0.32 & 0.75 & 2.35 \\
HD 199579 & 0.37 & 1.09 & 2.95 \\
HD 203938 & 0.74 & 2.15 & 2.91 \\
HD 206267 & 0.53 & 1.41 & 2.67 \\
HD 207198 & 0.62 & 1.50 & 2.42 \\
HD 207538 & 0.64 & 1.44 & 2.25 \\
HD 210839 & 0.57 & 1.58 & 2.78 \\
\enddata
\end{deluxetable}

\clearpage \clearpage

\begin{deluxetable}{cccc}
\tablecolumns{4}
\tablewidth{0pc}
\tablecaption{Absorption Lines Observed\label{linetable}}
\tablehead{\colhead{Wavelength (\AA{})} & \colhead{Oscillator Strength $f$} & \colhead{Damping Constant $\gamma$} & \colhead{Data Used}}
\startdata
951.0792 & $1.69\times10^{-4}$ & $1.30\times10^{8}$ & {\it FUSE} \\
951.2948 & $2.32\times10^{-5}$ & $1.29\times10^{8}$ & {\it FUSE} \\
953.4152 & $1.29\times10^{-2}$ & $2.19\times10^{8}$ & {\it FUSE} \\
953.6549 & $2.47\times10^{-2}$ & $2.10\times10^{8}$ & {\it FUSE} \\
959.4937 & $5.18\times10^{-5}$ & $1.18\times10^{8}$ & {\it FUSE} \\
1134.1563 & $1.46\times10^{-2}$ & $1.51\times10^{8}$ & {\it FUSE} \\
1134.4149 & $2.87\times10^{-2}$ & $1.49\times10^{8}$ & {\it FUSE} \\
1134.9803 & $4.16\times10^{-2}$ & $1.44\times10^{8}$ & {\it FUSE} \\
1159.8168 & $9.96\times10^{-6}$ & $4.70\times10^{8}$ & {\it FUSE} \\
1160.9366 & $2.75\times10^{-6}$ & $4.70\times10^{8}$ & {\it FUSE} \\
1199.5496 & $1.32\times10^{-1}$ & $4.07\times10^{8}$ & {\it HST} \\
1200.2233 & $8.69\times10^{-2}$ & $4.02\times10^{8}$ & {\it HST} \\
1200.7098 & $4.32\times10^{-2}$ & $4.00\times10^{8}$ & {\it HST} \\
\enddata
\end{deluxetable}

\clearpage \clearpage

\begin{deluxetable}{cccc}
\tablecolumns{4}
\tablewidth{0pc}
\tablecaption{Absorption Lines in the {\it FUSE} Wavelength Region Not Used in This Study\label{nolinetable}}
\tablehead{\colhead{Wavelength (\AA{})} & \colhead{Oscillator Strength $f$} & \colhead{Damping Constant $\gamma$} & \colhead{Reason Unused\tablenotemark{a}}}
\startdata
952.3034 & $2.29\times10^{-3}$ & $4.81\times10^{7}$ & $\Hmol$, OI \\
952.4148 & $1.97\times10^{-3}$ & $5.13\times10^{7}$ & $\Hmol$, OI \\
952.5227 & $5.18\times10^{-4}$ & $4.47\times10^{7}$ & $\Hmol$, OI \\
953.9699 & $3.31\times10^{-2}$ & $2.03\times10^{8}$ & $\Hmol$ \\
954.1042 & $4.00\times10^{-3}$ & $1.41\times10^{8}$ & $\Hmol$ \\
955.2644 & $6.92\times10^{-6}$ & $1.13\times10^{7}$ & $\Hmol$ \\
955.8816 & $1.91\times10^{-5}$ & $1.26\times10^{7}$ & $\Hmol$ \\
960.2014 & $1.17\times10^{-5}$ & $1.32\times10^{8}$ & Too weak\tablenotemark{b} \\
963.9903 & $1.24\times10^{-2}$ & $8.55\times10^{7}$ & $\Hmol$ \\
964.6256 & $7.90\times10^{-3}$ & $8.31\times10^{7}$ & $\Hmol$ \\
965.0413 & $3.86\times10^{-3}$ & $8.16\times10^{7}$ & $\Hmol$ \\
1003.3722 & $2.81\times10^{-8}$ & $3.99\times10^{8}$ & $\Hmol$, too weak\tablenotemark{c} \\
1003.3771 & $1.90\times10^{-7}$ & $3.99\times10^{8}$ & $\Hmol$, too weak\tablenotemark{c} \\
\enddata
\tablenotetext{a}{Describes the reason the line was not observed or conclusively measured, e.g. an atomic or $\Hmol$ line blends with the NI line or makes the continuum too uncertain to measure the NI line, or the NI line is too weak to observe in typical {\it FUSE} spectra.}
\tablenotetext{b}{The two lines of the 1160 \AA{} doublet have weaker oscillator strengths, but the S/N is characteristically better redward of 1100 \AA{}.  The weaker component of a doublet at 951 \AA{} has an oscillator strength nearly twice that of this line, and is only observed in five lines of sight.}
\tablenotetext{c}{Observing the 1003 \AA{} doublet would require a column density and/or S/N at least a combined order of magnitude larger than observed in our best cases.}
\end{deluxetable}

\clearpage \clearpage

\begin{deluxetable}{ccccccccccc}
\tablecolumns{11}
\tablewidth{0pc}
\tabletypesize{\tiny}
\rotate{}
\tablecaption{Measured Equivalent Widths:  {\it FUSE} Data\label{eqwidthsFUSE}}
\tablehead{\colhead{Sightline} & \colhead{$W_{951.1}$} & \colhead{$W_{951.3}$} & \colhead{$W_{953.4}$} & \colhead{$W_{953.7}$} & \colhead{$W_{959.5}$} & \colhead{$W_{1134.2}$} & \colhead{$W_{1134.4}$} & \colhead{$W_{1135.0}$} & \colhead{$W_{1159.8}$} & \colhead{$W_{1160.9}$} \\ \colhead{} & \colhead{(m\AA{})} & \colhead{(m\AA{})} & \colhead{(m\AA{})} & \colhead{(m\AA{})} & \colhead{(m\AA{})} & \colhead{(m\AA{})} & \colhead{(m\AA{})} & \colhead{(m\AA{})} & \colhead{(m\AA{})} & \colhead{(m\AA{})}}
\startdata
HD 24534 & \nodata & \nodata & $68.9\pm6.5$ & $83.9\pm7.0$ & \nodata & $99.6\pm4.4$ & $121.4\pm4.4$ & $136.5\pm4.4$ & $11.2\pm1.4$ & $5.7\pm2.1$ \\
HD 27778 & \nodata & \nodata & \nodata & \nodata & \nodata & $54.3\pm3.9$ & $62.5\pm3.6$ & $76.2\pm4.5$ & $13.2\pm3.2$ & $3.3\pm1.6$ \\
HD 37903 & \nodata & \nodata & $105.8\pm11.5$ & $116.3\pm11.8$ & \nodata & $161.3\pm2.9$ & $169.1\pm3.8$ & $174.8\pm3.3$ & $7.7\pm1.6$ & $3.8\pm3.2$ \\
HD 38087 & \nodata & \nodata & \nodata & \nodata & \nodata & $173.1\pm6.9$ & $183.7\pm6.6$ & $189.6\pm7.1$ & $7.8\pm4.0$ & \nodata \\
HD 40893 & \nodata & \nodata & $260.8\pm88.9$ & $294.1\pm126.1$ & $43.0\pm12.7$ & $293.1\pm13.6$ & $316.3\pm24.1$ & $332.9\pm31.1$ & $28.8\pm4.3$ & $4.4\pm3.1$ \\
HD 41117 & \nodata & \nodata & \nodata & \nodata & \nodata & $146.7\pm11.5$ & $170.3\pm5.1$ & $173.4\pm17.6$ & $32.8\pm19.5$ & \nodata \\
HD 42087 & \nodata & \nodata & $130.1\pm33.7$ & $100.7\pm28.8$ & $54.1\pm16.4$ & $148.7\pm3.5$ & $180.2\pm5.5$ & $250.8\pm9.2$ & $17.2\pm3.6$ & \nodata \\
HD 46056 & \nodata & \nodata & \nodata & \nodata & \nodata & $152.9\pm11.2$ & $165.9\pm11.3$ & $172.8\pm12.0$ & $25.3\pm9.9$ & \nodata \\
HD 46202 & $57.8\pm18.8$ & \nodata & $106.7\pm12.0$ & $118.9\pm13.8$ & $65.7\pm22.3$ & $138.1\pm14.4$ & $157.2\pm13.7$ & $166.8\pm14.2$ & \nodata & $7.8\pm3.5$ \\
HD 53367 & \nodata & \nodata & \nodata & \nodata & \nodata & $133.7\pm19.2$ & $159.0\pm17.3$ & $178.9\pm15.0$ & $24.4\pm4.3$ & \nodata \\
HD 73882 & \nodata & \nodata & \nodata & \nodata & \nodata & $191.7\pm6.7$ & $210.0\pm6.8$ & $219.1\pm6.3$ & $22.6\pm1.8$ & $8.2\pm2.1$ \\
HD 110432 & $30.0\pm3.8$ & $7.2\pm2.4$ & $80.5\pm3.9$ & $92.3\pm5.7$ & $23.2\pm4.0$ & $105.5\pm1.0$ & $118.1\pm1.1$ & $123.9\pm1.1$ & $14.3\pm1.5$ & $4.3\pm1.1$ \\
HD 147888 & \nodata & \nodata & $84.0\pm22.9$ & $78.7\pm26.5$ & $49.6\pm20.4$ & $116.5\pm5.0$ & $132.9\pm5.4$ & $124.8\pm13.2$ & $16.5\pm3.7$ & $5.8\pm3.5$ \\
HD 149404 & $62.8\pm5.0$ & $27.7\pm4.3$ & $131.7\pm8.7$ & $140.6\pm9.2$ & $42.9\pm4.2$ & $171.1\pm6.4$ & $179.7\pm6.5$ & $184.5\pm6.5$ & $34.1\pm1.0$ & $11.2\pm1.7$ \\
HD 152236 & \nodata & \nodata & \nodata & \nodata & \nodata & $296.0\pm27.3$ & $321.9\pm43.1$ & $340.6\pm49.5$ & $20.4\pm3.5$ & \nodata \\
HD 154368 & \nodata & \nodata & \nodata & \nodata & \nodata & $195.8\pm14.2$ & $206.3\pm16.8$ & $213.2\pm19.8$ & $41.7\pm7.8$ & \nodata \\
HD 167971 & \nodata & \nodata & \nodata & \nodata & \nodata & $223.8\pm23.6$ & $236.1\pm23.8$ & $242.7\pm23.9$ & $55.9\pm13.0$ & \nodata \\
HD 168076 & \nodata & \nodata & \nodata & \nodata & \nodata & $250.3\pm19.0$ & $277.7\pm17.2$ & $291.8\pm17.1$ & $38.7\pm8.9$ & \nodata \\
HD 170740 & \nodata & \nodata & $58.5\pm34.9$ & $85.2\pm24.4$ & $66.1\pm21.3$ & $97.8\pm3.7$ & $126.5\pm4.4$ & $123.0\pm4.5$ & \nodata & $10.0\pm4.0$ \\
HD 179406 & \nodata & \nodata & \nodata & \nodata & \nodata & $114.1\pm5.6$ & $121.6\pm5.4$ & $132.1\pm5.7$ & $16.6\pm10.6$ & \nodata \\
HD 185418 & \nodata & \nodata & $79.3\pm12.7$ & $95.0\pm14.7$ & \nodata & $107.6\pm1.1$ & $122.6\pm1.3$ & $133.4\pm1.4$ & $15.8\pm2.2$ & $2.6\pm2.1$ \\
HD 186994 & $22.1\pm11.5$ & \nodata & $224.8\pm12.2$ & $243.2\pm7.3$ & \nodata & $277.0\pm9.5$ & $289.8\pm10.2$ & $297.0\pm10.7$ & \nodata & \nodata \\
HD 192639 & $48.7\pm14.1$ & $38.0\pm15.0$ & $159.3\pm13.3$ & $178.4\pm14.6$ & $45.1\pm14.7$ & $246.4\pm7.0$ & $230.7\pm5.0$ & $251.0\pm6.8$ & $16.6\pm3.6$ & $4.8\pm3.4$ \\
HD 197512 & \nodata & \nodata & $142.3\pm16.8$ & $166.2\pm10.4$ & $76.5\pm17.8$ & $124.1\pm5.6$ & $139.7\pm6.2$ & $137.0\pm9.3$ & $6.4\pm1.9$ & \nodata \\
HD 199579 & $53.3\pm3.8$ & $15.6\pm3.5$ & $159.3\pm13.3$ & $178.4\pm14.6$ & $25.0\pm3.3$ & $179.6\pm2.0$ & $203.8\pm2.6$ & $210.2\pm3.1$ & $19.7\pm1.4$ & $6.6\pm2.2$ \\
HD 203938 & \nodata & \nodata & \nodata & \nodata & \nodata & $131.4\pm12.3$ & $139.9\pm12.1$ & $144.6\pm12.4$ & $29.1\pm7.8$ & \nodata \\
HD 206267 & $80.0\pm42.4$ & \nodata & $100.9\pm18.4$ & $129.1\pm10.7$ & $26.0\pm14.4$ & $156.3\pm2.4$ & $179.9\pm1.6$ & $159.0\pm6.4$ & $15.3\pm2.6$ & $6.3\pm2.4$ \\
HD 207198 & $62.7\pm24.2$ & \nodata & $126.6\pm33.5$ & $108.1\pm17.1$ & $63.4\pm21.0$ & $135.1\pm5.9$ & $142.6\pm5.8$ & $146.8\pm6.0$ & $22.7\pm2.6$ & $5.4\pm2.0$ \\
HD 207538 & \nodata & \nodata & $110.9\pm44.4$ & $125.6\pm30.4$ & \nodata & $125.3\pm6.4$ & $136.3\pm6.5$ & $141.9\pm6.2$ & $27.7\pm3.6$ & \nodata \\
HD 210839 & $55.3\pm4.8$ & $20.7\pm4.6$ & $133.2\pm8.6$ & $162.1\pm6.5$ & $22.4\pm3.8$ & $186.1\pm2.5$ & $193.2\pm3.0$ & $200.0\pm3.8$ & $18.6\pm1.1$ & $4.6\pm0.9$ \\
\enddata
\end{deluxetable}

\clearpage \clearpage

\begin{deluxetable}{cccc}
\tablecolumns{4}
\tablewidth{0pc}
\tablecaption{Measured Equivalent Widths:  {\it HST} Data\label{eqwidthsHST}}
\tablehead{\colhead{Sightline} & \colhead{$W_{1199.5}$} & \colhead{$W_{1200.2}$} & \colhead{$W_{1200.7}$} \\ \colhead{} & \colhead{(m\AA{})} & \colhead{(m\AA{})} & \colhead{(m\AA{})}}
\startdata
HD 27778 & $110.7\pm6.4$ & $103.5\pm7.1$ & $89.8\pm7.5$ \\
HD 37903 & $224.1\pm7.3$ & $215.6\pm8.0$ & $200.8\pm8.4$ \\
HD 147888 & $195.6\pm11.3$ & $182.5\pm11.4$ & $157.3\pm12.5$ \\
HD 185418 & $180.2\pm5.5$ & $170.4\pm6.1$ & $152.3\pm6.3$ \\
HD 192639 & $323.5\pm68.2$ & $303.3\pm49.2$ & $277.7\pm25.4$ \\
HD 206267 & $216.8\pm28.2$ & $201.4\pm20.6$ & $182.9\pm11.0$ \\
HD 207198 & $207.1\pm6.7$ & $198.9\pm6.9$ & $184.6\pm6.6$ \\
HD 207538 & $182.9\pm8.6$ & $172.7\pm8.4$ & $153.0\pm8.9$ \\
HD 210839 & $251.9\pm20.9$ & $235.0\pm15.4$ & $214.0\pm8.1$ \\
\enddata
\end{deluxetable}

\clearpage \clearpage
\begin{deluxetable}{cccccc}
\tablecolumns{6}
\tablewidth{0pc}
\tabletypesize{\scriptsize}
\tablecaption{Column Density Results\label{coldensities}}
\tablehead{\colhead{Sightline} & \colhead{log NI} & \colhead{$b$-value} & \colhead{log NH} & \colhead{N/H (ppm)} & \colhead{Method\tablenotemark{a}}}
\startdata
HD 24534 & $17.22^{+0.09}_{-0.08}$ & $4.9^{+0.9}_{-0.8}$ & $-4.12\pm0.09$ & $75^{+18}_{-14}$ & Uniform Error COG\tablenotemark{b} \\
HD 27778 & $17.11\pm0.16$ & $2.7\pm0.4$ & $-4.29^{+0.16}_{-0.18}$ & $51^{+23}_{-17}$ & COG \\
HD 37903 & $16.89^{+0.38}_{-0.35}$ & $8.8\pm0.8$ & $-4.61^{+0.38}_{-0.37}$ & $24^{+34}_{-14}$ & COG \\
HD 38087 & $16.83^{+0.31}_{-0.27}$ & $9.6\pm0.7$ & $-4.40^{+0.32}_{-0.49}$ & $40^{+42}_{-26}$ & COG \\
HD 40893 & $17.38^{+0.21}_{-0.19}$ & $15.4^{+1.9}_{-1.8}$ & $-4.21^{+0.22}_{-0.23}$ & $61^{+39}_{-25}$ & COG \\
HD 41117 & $18.53^{+0.16}_{-1.38}$ & $4.6^{+3.6}_{-2.0}$ & $-3.01^{+0.18}_{-2.99}$ & $976^{+484}_{-975}$ & COG \\
HD 42087 & $17.23^{+0.14}_{-0.12}$ & $8.3^{+2.1}_{-1.8}$ & $-4.26^{+0.15}_{-0.18}$ & $54^{+23}_{-18}$ & Uniform Error COG \\
HD 46056 & $17.42^{+0.27}_{-0.23}$ & $7.5\pm0.6$ & $-4.11^{+0.28}_{-0.30}$ & $78^{+69}_{-38}$ & COG \\
HD 46202 & $17.42^{+0.46}_{-0.37}$ & $6.9\pm1.2$ & $-4.26^{+0.46}_{-0.47}$ & $55^{+104}_{-36}$ & COG \\
HD 53367 & $17.40^{+0.07}_{-0.06}$ & $7.2\pm0.5$ & $-4.23^{+0.12}_{-0.31}$ & $58^{+17}_{-29}$ & COG\tablenotemark{b} \\
HD 73882 & $17.36\pm0.11$ & $9.8^{+0.6}_{-0.5}$ & $-4.23^{+0.12}_{-0.15}$ & $59^{+19}_{-17}$ & COG \\
HD 110432 & $17.09^{+0.19}_{-0.18}$ & $5.5\pm0.3$ & $-4.11^{+0.19}_{-0.21}$ & $77^{+44}_{-30}$ & COG \\
HD 147888 & $17.38^{+0.34}_{-0.47}$ & $5.8^{+1.2}_{-1.0}$ & $-4.38^{+0.34}_{-0.51}$ & $41^{+50}_{-29}$ & COG \\
HD 149404 & $17.56^{+0.06}_{-0.05}$ & $7.8\pm0.6$ & $-4.01^{+0.10}_{-0.15}$ & $96^{+23}_{-28}$ & COG \\
HD 152236 & $17.27\pm0.08$ & $15.9\pm1.4$ & $-4.57^{+0.11}_{-0.18}$ & $26^{+8}_{-9}$ & COG \\
HD 154368 & $17.67^{+0.13}_{-0.11}$ & $9.2^{+0.7}_{-0.6}$ & $-3.92^{+0.14}_{-0.13}$ & $120^{+44}_{-31}$ & COG \\
HD 167971 & $17.83^{+0.18}_{-0.15}$ & $10.3^{+1.1}_{-0.9}$ & $-3.90^{+0.21}_{-0.69}$ & $125^{+79}_{- 99}$ & COG \\
HD 168076 & $17.59^{+0.13}_{-0.12}$ & $12.7\pm0.8$ & $-4.14^{+0.17}_{-0.43}$ & $71^{+34}_{-44}$ & COG \\
HD 170740 & $17.65^{+0.56}_{-0.37}$ & $4.8^{+0.6}_{-1.5}$ & $-3.76^{+0.56}_{-0.41}$ & $175^{+462}_{-107}$ & COG \\
HD 179406 & $17.27^{+1.00}_{-0.77}$ & $5.6^{+1.1}_{-2.2}$ & $-4.17^{+1.00}_{-0.89}$ & $67^{+604}_{-58}$ & COG \\
HD 185418 & $17.04^{+0.25}_{-0.23}$ & $5.9^{+2.0}_{-1.6}$ & $-4.35^{+0.25}_{-0.27}$ & $44^{+35}_{-21}$ & Uniform Error COG \\
HD 186994 & $16.59^{+1.03}_{-0.52}$ & $16.8^{+2.9}_{-3.4}$ & $-4.35^{+1.03}_{-0.68}$ & $44^{+433}_{-35}$ & COG \\
HD 192639 & $17.18^{+0.48}_{-0.40}$ & $12.0\pm1.4$ & $-4.31^{+0.48}_{-0.44}$ & $49^{+ 99}_{-31}$ & COG \\
HD 197512 & $16.96^{+0.21}_{-0.19}$ & $7.3\pm0.6$ & $-4.48^{+0.22}_{-0.26}$ & $33^{+21}_{-15}$ & COG \\
HD 199579 & $17.16\pm0.10$ & $9.7^{+0.4}_{-0.3}$ & $-4.09^{+0.11}_{-0.14}$ & $81^{+24}_{-22}$ & COG \\
HD 203938 & $17.53^{+0.19}_{-0.17}$ & $6.1\pm0.6$ & $-4.17^{+0.20}_{-0.23}$ & $67^{+39}_{-27}$ & COG \\
HD 206267 & $17.16^{+0.27}_{-0.29}$ & $8.5\pm0.6$ & $-4.38^{+0.27}_{-0.34}$ & $41^{+36}_{-22}$ & COG \\
HD 207198 & $17.41^{+0.18}_{-0.19}$ & $6.7\pm0.7$ & $-4.14^{+0.19}_{-0.27}$ & $72^{+40}_{-33}$ & COG \\
HD 207538 & $17.39\pm0.17$ & $6.0\pm0.6$ & $-4.19^{+0.18}_{-0.21}$ & $64^{+32}_{-24}$ & COG \\
HD 210839 & $17.19^{+0.13}_{-0.12}$ & $9.5\pm0.5$ & $-4.26\pm0.14$ & $55^{+20}_{-15}$ & COG \\
\enddata
\tablenotetext{a}{Methods:  COG---Standard curve-of-growth method.  Uniform Error COG---Curve-of-growth method with fractional errors averaged.}
\tablenotetext{b}{This solution is taken from a local minimum in the $\chi^2$ array (from the stated curve-of-growth method) that is not the global minimum; the global minimum is ruled out based on inconsistency with the line profiles.  See \S\ref{ss:alternate} for further discussion.}
\end{deluxetable}

\clearpage \clearpage


\begin{thebibliography}{}


\bibitem[Bohlin et al.(1983)]{Bohlin83} Bohlin, R. C. et al.  1983, ApJS, 51, 277.

\bibitem[Bohlin et al.(1978)]{BohlinSavDrake} Bohlin, R. C., Savage, B. D., \& Drake, J. F.  1978, ApJ, 224, 132.

\bibitem[Butler \& Dalgarno(1979)]{ButlerDalgarno} Butler, S. E. \& Dalgarno, A.  1979, ApJ, 234, 765.


\bibitem[Cartledge et al.(2004)]{Cartledge2} Cartledge, S. I. B., Lauroesch, J. T., Meyer, D. M., \& Sofia, U. J.  2004, ApJ, 613, 1037.

\bibitem[Cartledge et al.(2006)]{Cartledge3} Cartledge, S. I. B., Lauroesch, J. T., Meyer, D. M., \& Sofia, U. J.  2006, ApJ, 641, 327.

\bibitem[Cowan et al.(1982)Cowan, Hobbs, \& York]{Cowan} Cowan, R. D., Hobbs, L. M., \& York, D. G.  1982, ApJ, 257, 373.


\bibitem[Diplas \& Savage(1994)]{Diplas} Diplas, A. \& Savage, B. D.  1994, ApJ, 93, 211.

\bibitem[Feldman et al.(2001)]{Feldman} Feldman, P. D., Shanow, D. J., Kruk, J. W., Murphy, E. M., \& Moos, H. W.  2001, JGR, 106, 8119.

\bibitem[Ferlet(1981)]{Ferlet} Ferlet, R.  1981, A\&A, 98, L1.


\bibitem[Fitzpatrick \& Massa(1990)]{FitzMassa} Fitzpatrick, E. L. \& Massa, D.  1990, ApJS, 72, 163.

\bibitem[Grevesse \& Noels(1993)]{Grevesse} Grevesse, N., \& Noels, A.  1993, {\it Origin and Evolution of the Elements}, ed. N. Prantzos, E. Vangioni-Flam, \& M. Casse (Cambridge:  Cambridge Univ. Press), 15.

\bibitem[Hanson et al.(1992)]{Hanson} Hanson, M. M., Snow, T. P., \& Black, J. H.  1992, ApJ, 392, 571.

\bibitem[Hibbert et al.(1985)Hibbert, Dufton, \& Keenan]{Hibbert} Hibbert, A., Dufton, P. L., \& Keenan, F. P.  1985, MNRAS, 213, 721.

\bibitem[Holweger(2001)]{Holweger} Holweger, H.  2001, AIP Conf. Proc. 598: Joint SOHO/ACE workshop ``Solar and Galactic Composition'', 23.

\bibitem[Jenkins et al.(2000)]{Jenkins} Jenkins, E. B. et al.  2000, ApJ, 538, L81.

\bibitem[Jenkins et al.(1999)]{Jenkins1999}  Jenkins, E. B., Trip, T. M., Wozniak, P. R., \& Sofia, U. J.  1999, ApJ, 520, 182.


\bibitem[Jensen et al.(2005)Jensen, Rachford, \& Snow]{Jensen} Jensen, A. G., Rachford, B. L., \& Snow, T. P.  2005, ApJ, 619, 891.


\bibitem[Knauth et al.(2003)]{Knauth} Knauth, D. C., Andersson, B.-G., McCandliss, S. R., \& Moos, H. W.  2003, ApJ, 596, L51.


\bibitem[Lodders(2003)]{Lodders} Lodders, K.  2003, ApJ, 591, 1220.

\bibitem[Lugger et al.(1978)]{Lugger} Lugger, P. M., York, D. G., Blanchard, T. \& Morton, D. C.  1978, ApJ, 224, 1059.


\bibitem[Meyer et al.(1997)Meyer, Cardelli, \& Sofia]{Meyer} Meyer, D. M., Cardelli, J. A. \& Sofia, U. J.  1997, ApJ, 490, L103.


\bibitem[Micol et al.(1999)]{Micol} Micol, A., Durand, D., Pirenne, B., Gaudet, S., \& Hodge, P.  1999, ASP Conference Series, Vol. 172, ``Astronomical Data Analysis Software and Systems VIII'', 191.

\bibitem[Moos et al.(2002)]{Moos} Moos, H. W. et al.  2002, ApJS, 140, 3.


\bibitem[Morton(2003)]{Morton03} Morton, D. C.  2003, ApJS, 149, 205.


\bibitem[Pan et al.(2004)]{Pan} Pan, K., Federman, S. R., Cunha, K., Smith, V. V., \& Welty, D. E.  2004, ApJS, 151, 313.

\bibitem[Rachford et al.(2001)]{Rachford110432} Rachford, B. L. et al.  2001, ApJ, 555, 839.

\bibitem[Rachford et al.(2002)]{Rachford} Rachford, B. L. et al.  2002, ApJ, 577, 221.



\bibitem[Shull \& Van Steenburg(1985)]{ShullVS} Shull, J. M. \& Van Steenburg, M.  1985, ApJ, 294, 599.

\bibitem[Snow(2000)]{SnowJGR} Snow, T. P.  2000, JGR, 105, 10239.

\bibitem[Snow et al.(1996)]{Snow154368} Snow, T. P., Black, J. H., van Dishoeck, E. F., Burks, G., Crutcher, R. M., Lutz, B. L., Hanson, M. M., \& Shuping, R. Y.  1996, ApJ, 465, 245.


\bibitem[Snow \& Witt(1996)]{SnowWitt} Snow, T. P. \& Witt, A. N.  1996, ApJ, 468, L65.


\bibitem[Sofia et al.(1994)Sofia, Cardelli, \& Savage]{SofiaCS} Sofia, U. J, Cardelli, J. A., \& Savage, B. D.  1994, ApJ, 430, 650.

\bibitem[Sofia \& Meyer(2001)]{SofiaMeyer} Sofia, U. J. \& Meyer, D. M.  2001, ApJ, 554, L221.

\bibitem[Sonneborn et al.(2000)]{Sonneborn}  Sonneborn, G. et al.  2000, ApJ, 545, 277.


\bibitem[Tachiev \& Froese Fischer(2002)]{Tachiev} Tachiev, G. I. \& Froese Fischer, C.  2002, A\&A, 385, 716.

\bibitem[Walborn(1976)]{Walborn} Walborn, N. R.  1976, ApJ, 205, 419.

\bibitem[Welty \& Fowler(1992)]{WeltyFowler} Welty, D. E. \& Fowler, J. R.  1992, ApJ, 393, 193.

\bibitem[Welty \& Hobbs(2001)]{WeltyKI} Welty, D. E. \& Hobbs, L. M..  2001, ApJS, 133, 345.

\bibitem[Welty et al.(2003)Welty, Hobbs, \& Morton]{WeltyCaI} Welty, D. E., Hobbs, L. M., \& Morton, D. C.  2003, ApJS, 147, 61.


\bibitem[Withbroe(1971)]{Withbroe} Withbroe, G. K.  1971, {\it The Menzel Symposium}, ed. K. B. Gebbie (NBS Spec. Pub. 353; Washington, D.C.:  GPO).

\bibitem[York et al.(1983)]{York} York, D. G., Spitzer, L., Bohlin, R. C., Hill, J., Jenkins, E. B., Savage, B. D., \& Snow, T. P.  1983, ApJ, 266, L55.


\end{thebibliography}
\end{document}